\documentstyle[aps,epsfig,amsmath,epsf]{revtex}
\newcommand{\cd}{\makebox[0.08cm]{$\cdot$}}
\newcommand{\bg}[1]{\mbox{\boldmath $#1$}}
\newcommand{\sla}{\not\!}

\begin{document}
\title{ Renormalized nonperturbative fermion model \\ in Covariant Light Front
Dynamics}
\author{V.A. Karmanov \thanks{e-mail: karmanov@sci.lebedev.ru,
karmanov@lpsc.in2p3.fr}}
\address{Lebedev Physical Institute, Leninsky Prospekt 53, 119991
Moscow, Russia}
\author{J.-F. Mathiot \thanks{e-mail: mathiot@in2p3.fr}}
\address{Laboratoire de Physique Corpusculaire,
Universit\'e Blaise-Pascal,\\ CNRS/IN2P3, 24 avenue des
Landais, F-63177 Aubi\`ere Cedex, France}
\author{A.V. Smirnov \thanks{e-mail: asmirnov@sci.lebedev.ru}}
\address{Lebedev Physical Institute, Leninsky Prospekt 53, 119991
Moscow, Russia}
\maketitle

\begin{abstract}
Within the framework of the covariant formulation of Light-Front
Dynamics,
we develop a nonperturbative renormalization scheme in the fermion model 
supposing that the composite fermion is a superposition  of the "bare"
fermion and a fermion+boson state. We first assume the constituent
boson to be spinless. Then we address the case of gauge bosons in the Feynman and 
in the Light-Cone gauges.
For all these cases the fermion state vector and the necessary renormalization counterterms
are calculated analytically. It turns out that in Light-Front Dynamics, 
to restore the rotational invariance, an extra counterterm is
needed, having no any analogue in Feynman approach.  For gauge bosons the results obtained in the two 
gauges are compared with each other. In
general, the number of spin components of the two-body (fermion+boson) wave function depends
on the gauge. But due to the two-body Fock space truncation, only one non-zero
component survives for each gauge. And moreover, the whole solutions for the
state vector, found for the Feynman and Light-Cone gauges, are the same
(except for the normalization factor). The counterterms are however different.
\end{abstract}
\newpage

\bibliographystyle{unsrt}

\section{Introduction}

The understanding of hadronic systems is at the heart of numerous
theoretical studies these last 20 years. While the description of
experimental processes at very high energy can rely on perturbation
theory, the study of bound states necessitates the use of
non-per\-tur\-ba\-tive methods.
Concerning the latters, the nonperturbative renormalization procedure and the
invariance of the system under gauge transformations are the properties of
particular interest.
Among the various frameworks
used over the past to tackle these problems, we shall concentrate
on Light Front Dynamics (LFD)~\cite{dirac} which exists in two
field-theoretical forms: standard LFD~\cite{bpp} and 
explicitly Covariant Light Front
Dynamics (CLFD)~\cite{cdkm}. Standard LFD deals with the
state vector defined on the plane $t+z=0$, while in CLFD this plane 
is given by the invariant equation $\omega\cd x=0$, where $\omega$ is an
arbitrary four-vector with $\omega^2=0$.  The particular choice 
$\omega=(1,0,0,-1)$ turns CLFD into standard LFD.
We shall see below that the explicit covariance of CLFD is a very
powerfull tool to address the above questions in a transparent way.

Following the pioneering work of Wilson~\cite{wils65}, the perturbative
renormalization in QED has been studied both in
LFD~\cite{bro98,must91,lb95} and CLFD~\cite{dugne}. In these papers 
the fermion self-energy is investigated, and the necessary counterterms
needed to renormalize it are found. To generalize the renormalization 
technique for
nonperturbative processes, it is necessary to consider the same
questions
in terms of the Fock decomposition of the state vector. This has been
already done for scalar
models in LFD~\cite{perr90,ji,hara97} and CLFD~\cite{bernard}. 
Such toy
models are important to settle general equations and test
numerical procedures. At the same time they are too simple because of,
firstly, their super-renormalizability, and, secondly, the absense of contact
interactions appearing for non-zero spin particles in LFD and CLFD.
The next step therefore is an investigation of
fermionic composite states~\cite{perr90,perr91,glp,brs73,bhc}.
Two important properties have emerged from these calculations.

{\it i)} The possible Fock sector dependence of the (mass) counterterm,
which has been discussed in Ref.~\cite{perr90} for LFD. The necessity to
remove, by hands, the counterterm in the equation which defines the last
Fock state component considered in the calculation is now well
understood~\cite{hara97,bernard}. This (infinite) counterterm should be
balanced by a (also infinite) contribution arising from higher Fock state
components not taken into account due to the truncation of the Fock state
expansion. 

{\it ii)} The appearance of non-local counterterms needed
to recover known symmetries and to renormalize the theory~\cite{must91}.
As has already been shown in CLFD~\cite{dugne}, the
knowledge of the structure of the counterterms as functions of the
light front plane orientation
is necessary to understand how non-local counterterms appear in
LFD. We shall come back extensively to this question in the following.

We consider in the present study a fermion $F$ coupled to
bosons $b$ with spin either zero or one. We  restrict
ourselves to the approximation where only the first two Fock sectors
($|F\rangle + |Fb\rangle$)
in the state vector are retained. Such a model is instructive in many aspects:
{(\it i)} since we are dealing with fermions, we have to take into
account contact interactions. This, also, implies non-trivial
contributions arising from the counterterms needed to renormalize the
theory;
{(\it ii)} we can investigate the case of gauge theories;
{(\it iii)} the procedure can be checked by comparison with results of
previous calculations based
on perturbative methods, as mentioned above. It can thus be extended with
some confidence to calculations involving higher Fock state components.

In a first step, we suppose the boson $b$ to be spinless, that just 
corresponds to the well-known Yukawa model. This gives
us an opportunity to develop our method and show the peculiarities of
CLFD. We reproduce the results
obtained in this model in Ref.~\cite{glp} in standard LFD.

We then study the system where the constituent boson is a gauge
particle. In order to investigate the influence of the gauge on
the state vector, we consider two different gauges: the
Feynman and the Light-Cone (LC) ones. We will see that the
number of spin components of the two-body wave function depends on
the gauge. In the Feynman gauge the wave function is determined, in
general, by eight components, whereas in the LC gauge it 
contains four components. However, in the simple model
developed in the present paper where we restrict ourselves to the
$|F\rangle+|Fb\rangle$ Tamm-Dancoff truncation, 
three non-zero components remain for the Feynman gauge and two ones for 
the LC gauge. Moreover, after imposing a certain condition on the
counterterms, only one non-zero component of the two-body 
wave function survives for each gauge.

The plan of the paper is as follows. In Sec.~\ref{general} we present
our general framework. In Sec.~\ref{need} the need for 
a specific extra counterterm in LFD is justified.
In Sec.~\ref{scalar} we calculate the state vector and the counterterms 
for the scalar boson case. 
In Sec.~\ref{vec_feyn} we solve the problem for the gauge
boson in the Feynman gauge. The same problem, but for the LC gauge, is
solved in Sec.~\ref{vec_lc}. The results are summarized in 
Sec.\ref{discuss}, where we discuss also the questions of gauge invariance and
comparison with perturbative calculations. Sec.~\ref{concl} contains our
concluding remarks. Technical derivations are rounded up in Appendices.


\section{General Considerations}\label{general}


\subsection{Exact Lagrangian for scalar bosons}
In order to clearly settle the starting point of our study, 
we shall recall in this section some known results. 
For simplicity, we detail only the scalar boson case. 
The case of vector bosons can be deduced very easily.

We shall start with the  bare  Lagrangian written as
\begin{equation}
\label{eq1b}
{\cal L}=\bar{\Psi}_{0}\left[i{\sla \partial} -
m_{0}\right] \Psi_{0} +\frac{1}{2}\left[\partial_{\nu}\Phi_{0}
\partial^{\nu}\Phi_{0}-\mu_{0}^2\Phi_{0}^2\right] +
g_0\bar{\Psi}_0\Psi_0 \Phi_0,
\end{equation}
where ${\sla \partial}=\gamma\cd\partial\equiv
\gamma^{\nu}\partial_{\nu}$, 
$\Psi_0$ ($\Phi_0$) is the bare fermion (boson) field,
$m_0$ ($\mu_0$) is the bare fermion (boson) mass, and $g_0$ is
the bare coupling constant.
Going over from bare to dressed quantities by means of the standard
substitutions
\begin{equation}
\label{eq2b}
\Psi_0=Z_2^{1/2}\Psi,\,\,\,\,\Phi_0=Z_3^{1/2}\Phi,\,\,\,\,
gZ_1=g_0Z_2Z_3^{1/2},
\end{equation}
where $g$ is the physically observed value of the coupling constant,
and introducing also the physical masses $m$ and $\mu$, we get
\begin{multline}
\label{eq3bb}
{\cal L}=\bar{\Psi}\left[i{\sla \partial}-
m\right] \Psi +\frac{1}{2}\left[\partial_{\nu}\Phi
\partial^{\nu}\Phi-\mu^2\Phi^2\right] +
g\bar{\Psi}\Psi \Phi  \\
+(Z_{2}-1)\bar{\Psi}\left[i{\sla \partial}-
m\right] \Psi +\frac{(Z_{3}-1)}{2}\left[\partial_{\nu}\Phi
\partial^{\nu}\Phi-\mu^2\Phi^2\right]\\
+\delta m\bar{\Psi}\Psi+
\frac{1}{2}\delta \mu^{2} \Phi^2+g(Z_1-1)\bar{\Psi}\Psi \Phi.
\end{multline}
From Eqs.~(\ref{eq1b})--(\ref{eq3bb}) it is easy to find that
\begin{equation}
\label{dmdmu}
\delta m=Z_2(m-m_0),\quad \delta \mu^{2}=Z_3(\mu^{2}-\mu_0^{2}).
\end{equation}

The counterterms $Z_{1-3}$, $\delta m$, and $\delta \mu^{2}$ can be found 
from the
definitions of the observed quantities $g$, $m$, and $\mu^{2}$.
Namely, $\delta m$ and $\delta\mu^2$ are defined by the positions
of the one-fermion and one-boson propagator poles.  
The additional requirement that the residues of these propagators
at their poles equal $i$ allows to fix the constants $Z_2$ and $Z_3$.
Finally, the constant $Z_1$ is determined from the condition of the 
coincidence of the on-shell dressed fermion-fermion-boson vertex at zero boson four-momentum
with the physical coupling constant.

\subsection{Perturbative vs Fock state decomposition}\label{PFS}

As we already mentioned in the Introduction, 
a natural framework to study composite 
systems in LFD is the Fock state decomposition. Renormalization 
procedures,
on the other hand, are well under control in perturbative Quantum Field 
Theory
using the 4-Dimensional Feynman (4DF) diagrammatic expansion. In the latter
approach, the composite fermion mass is defined as the pole of the fermion 
2-Point Green's Function (2PGF) which is obtained
by summing up the chain series of the iterations of the
fermion self-energy (a sum of
all irreducible Feynman diagrams calculated up to a given order
of coupling constant), $\Sigma_{4DF}(\sla{p},m_0)$, according to
\begin{equation}
\frac{i}{\sla p -m_{0}}+ \frac{i}{\sla{p} -m_{0}}
\left[-i\Sigma_{4DF}(\sla p, m_{0})\right]\frac{i}{\sla p -m_{0}}+\ldots\
=\frac{i}{\sla{p}-m_0-\Sigma_{4DF}(\sla{p},m_0)}.
\label{freeprop}
\end{equation}

In LFD state vectors are defined as eigenstates of the corresponding
light-front Hamiltonians. By using the Fock decomposition of the state vector,
the eigenstate problem can be easily reduced
to that of solving a system of linear homogeneous equations
which involve various Fock components. Note that the Fock decomposition 
is made in terms of the
number of Fock components rather than the coupling constant powers. 
Each Fock component effectively brings 
irreducible contributions of order $\alpha$ or higher, with
$\alpha=\frac{g^{2}}{4\pi}$, to the fermion propagator.
The summation of these contributions to all
orders in $\alpha$ is done implicitly when solving the system of linear
equations mentioned above. 
The  mass of the composite particle is found from the condition that
this system has a nontrivial solution. 
Such a procedure is an analogue, in LFD, 
of the summation of the chain series for the 2PGF in 4DF approach. 
Below we shall derive and solve this
system of equations, for the Fock decomposition restricted to the
components $|F\rangle+|Fb\rangle$, for scalar and gauge bosons in
both the Feynman and LC gauges.

\subsection{Renormalization scheme} \label{RS}
Before going into more detailed and technical considerations, it is
necessary to clarify in which scheme we shall work to renormalize the
system under study. A nice exposition of the standard derivation can be
found in~\cite{peskin,iz}. The following two schemes are usually
considered in perturbative calculations:
\begin{itemize}
\item {\it Bare renormalization scheme}. The starting point is the bare 
Lagrangian
with bare masses and coupling constants, as given by Eq.~(\ref{eq1b}). 
These masses and coupling
constants are determined from physical conditions. For
instance, the bare fermion mass is chosen so to get a pole of the 2PGF at the 
physical fermion mass. The bare mass expressed in terms of the physical one 
is in general infinite and must be regularized.
\item {\it Dressed renormalization scheme}. The starting point here is a
renormalized Lagrangian~(\ref{eq3bb}) defined in terms of (finite) physical 
masses and coupling
constants, and including the counterterms which 
are thus infinite and must be regularized.
\end{itemize}
The two regularization schemes should of course give the same 
physical observables provided no
approximations have been made. The latter never happens in practice: 
one either makes a
perturbative expansion, or truncate Fock space in LFD, or both. 
The equivalence between the two schemes 
can be directly checked in the lowest order of perturbation theory, as 
explained in any text book on Quantum Field Theory. In LFD, this 
question was partially adressed in Refs.~\cite{wils65,glp}. We 
have also tackled it within CLFD using perturbative 
expansions~\cite{dugne}.

Nonperturbative calculations however hardly admit such a check. In the 
spirit of the Fock state decomposition, one should make sure that the 
one-particle states are as close as possible to the 
physical ones. In our case this implies that the basic one-fermion
state $|F\rangle$ and the full state $|F\rangle+|Fb\rangle$
correspond to the same physical mass $m$,
and we thus should consider 
the mass counterterm $\delta m$. 

The counterterm proportional to $Z_2-1$ in Eq.~(\ref{eq3bb})
renormalizes the fermion field.
However, since we are working here with the state vector
rather than with particle fields, it is
more appropriate to deal directly with the state vector normalization factor
$N$ instead of $Z_2$. For this reason, we will exclude the counterterm 
with $Z_2$ from our consideration. The normalization factor $N$
does not enter the
eigenstate equation for the state vector and can be calculated after the solution
is found. 

Our task is simplified for the Fock decomposition truncated 
to the components $|F\rangle+|Fb\rangle$. In this case, the boson state 
vector is still free, therefore, $Z_3=1$ and $\delta \mu^2=0$.

Finally, the counterterm with $Z_1$ is related to the coupling constant 
renormalization. Since in the present paper we do not calculate  
physical observables like electromagnetic form factors or  
scattering amplitudes, we may deal with the 
unrenormalized coupling constant and put $Z_1=1$. For convenience, we 
shall call this coupling constant $g$.

So, as far as the counterterms in the Lagrangian are concerned 
[the second and the third lines in Eq.~(\ref{eq3bb})], it is enough for solving the 
nonperturbative problem in the 
two-body approximation, to retain explicitly the term 
with $\delta m$ only, while those containing 
$Z_1$, $Z_2$, $Z_3$, and $\delta \mu^2$ can be dropped out. The quantity
$\delta m$ defined now as
\begin{equation}
\label{mm0}
\delta m=m-m_0,
\end{equation}
that is a shift of mass caused by the interaction.

\section{The need for a new counterterm in LFD} \label{need}
As we shall see explicitly in the following applications, 
it turns out that one has to introduce in LFD another
counterterm which is not reduced to any of those considered above.

To explain its role, let us calculate the light-front 2PGF in the 
dressed renormalization scheme, using the two-body approximation. 
Below in this section we obtain a formal
expression for the 2PGF, then find the perturbative
(up to terms of order $g^4$) fermion-boson on-shell scattering amplitude 
$M^{(4)}$ and
demonstrate that without an additional counterterm such an
amplitude would be dependent on the light front orientation 
(i.~e. on the four-vector $\omega$). The dependence of approximate
physical CLFD amplitudes on $\omega$ is equivalent to the rotational
symmetry violation in standard LFD.

Since we are not dealing here with an eigenstate problem, we should
retain the constant $Z_2$ in the Lagrangian~(\ref{eq3bb}), while
putting $Z_1=Z_3=1$, $\delta \mu^2=0$, as explained above. 
So, the Lagrangian we consider in this section is
\begin{equation}
\label{eq3bba}
{\cal L}=\bar{\Psi}\left[i\sla{\partial}-
m\right] \Psi +\frac{1}{2}\left[\partial_{\nu}\Phi
\partial^{\nu}\Phi-\mu^2\Phi^2\right]
+g\bar{\Psi}\Psi \Phi+(Z_{2}-1)\bar{\Psi}
\left[i\sla{\partial}-m\right] \Psi
+\delta m\bar{\Psi}\Psi.
\end{equation}
In the following we will use the fact that perturbative expansions
of the quantities $Z_2-1$ and $\delta m$ start with terms of order $g^2$.

\begin{figure}[btph]
\begin{center}
\centerline{\epsfbox{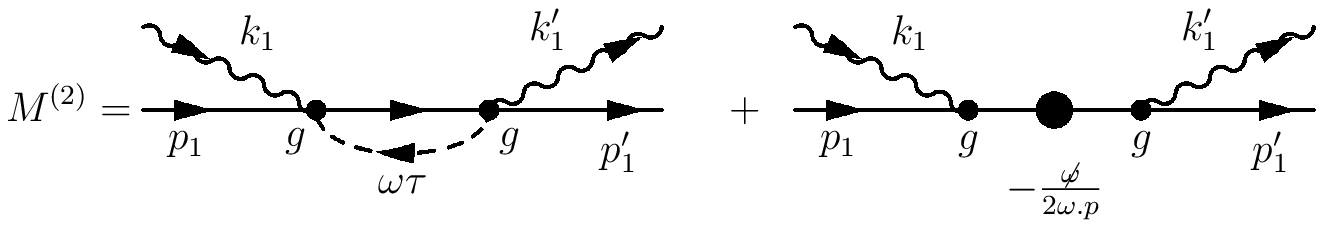}}
\caption{Fermion-boson CLFD scattering amplitude in $g^2$-order
of perturbation theory}.\label{fig0}
\end{center}
\end{figure}

Denote by $p_1$ ($p_1'$) the initial (final) fermion four-momentum;
the corresponding boson four-momenta are $k_1$ ($k_1'$).
In the lowest (second) order in $g$ the on-energy-shell 
scattering amplitude $M^{(2)}$
is defined by the sum of the two CLFD diagrams shown in Fig.~\ref{fig0}. 
The rules of the CLFD diagram technique are exposed in detail in 
Ref.~\cite{cdkm}.
As usual, straight solid, wavy, and dashed lines correspond to fermions,
bosons, and spurions, respectively. The internal fermion line
with a large full blob in the second diagram is the so-called
contact term\footnote{In Ref.~\cite{cdkm} the contact terms were represented
by lines with crosses. Here we changed this notation, reserving
the cross symbol for counterterms (see below).}. The amplitude 
$M^{(2)}$ was calculated in Ref.~\cite{cdkm}:
\begin{equation}
\label{amplm2}
M^{(2)}=g^2\bar{u}(p_1')\left[\frac{\sla{p}+\sla{\omega}\tau+m}
{m^2-p^2}-\frac{\sla{\omega}}{2(\omega\cd p)}
\right]u(p_1),
\end{equation}
where $p=p_1+k_1=p'_1+k'_1$, $\tau=(m^2-p^2)/2(\omega\cd p)$. The two addendums
in the square brackets on the r.~h.~s. of Eq.~(\ref{amplm2}) just correspond 
to the contributions from the two diagrams shown in Fig.~\ref{fig0}. It is easy
to see that the $\omega$-dependent terms in the sum cancel each other,
the whole amplitude $M^{(2)}$ being $\omega$-independent:
\begin{equation}
\label{amplm2a}
M^{(2)}=ig^2\bar{u}(p_1')D(p)u(p_1),
\end{equation}
where
\begin{equation}
\label{d2}
D(p)=i\frac{\sla{p}+m}{p^2-m^2}
\end{equation}
is the 2PGF in the lowest (zeroth) order of perturbation theory. Note that it
has the same form as in the 4DF approach, as it should.

The exact light-front 
2PGF can be obtained by formal summing up the perturbation series.
Such a procedure is quite similar to that used
in standard QED and leading to the Dyson equation. The difference is
that LFD calculations involve contact terms mentioned above.

Let us define the mass operator $-{\cal M}(p)$ (the sign minus 
is introduced for convenience) which is
a sum of all irreducible light-front diagrams containing no
one-fermion intermediate states or, in other words, including only 
intermediate states with bosons. 
The equation for the exact 2PGF ${\cal D}(p)$ 
is obtained by infinite iterations of the kernel being a sum of the three terms:
the vertices $(Z_2-1)(\sla{p}-m)$, $\delta m$, and the mass operator
$-{\cal M}(p)$. This equation reads
\begin{equation}
\label{2PGFexact}
{\cal D}(p)=D(p)+{\cal D}(p)[i(Z_2-1)(\sla{p}-m)+i\delta m-i{\cal M}(p)]D(p).
\end{equation}
It has the formal solution
\begin{multline}
\label{solD}
{\cal D}(p)=D(p)
\frac{1}
{1-i
[(Z_2-1)(\sla{p}-m)+
\delta m-
{\cal M}(p)]
D(p)}\\
=
i\frac{\sla{p}+m}
{p^2-m^2+(Z_2-1)(p^2-m^2)+
[\delta m-{\cal M}(p)]
(\sla{p}+m)}.
\end{multline}

In the $g^2$ (or two-body) approximation the mass operator 
is given by the sum of the two diagrams shown in Fig.~\ref{fig0a}. 
The diagrams (a) and (b) taken with the minus sign each are the
light-front fermion self-energy, $-\Sigma(p)$, and a contact
term on the fermion line, $-\Sigma_{fc}(p)$, respectively. 
Note that the amplitudes of these diagrams include integrals
over bosonic momentum. Because of different phase space volumes, 
it is not judicious to combine the contact term
with the usual light-front fermion propagator in the self-energy
diagram into the unique term $D(p)$, like it has been done in 
Eqs.~(\ref{amplm2})--(\ref{d2}). So, the diagrams (a) and (b)
in Fig.~\ref{fig0a} must be treated separately.
The mass operator ${\cal M}(p)$ is a light-front analogue of the 4DF 
fermion self-energy $\Sigma_{4DF}(\sla{p},m)$ calculated in the
dressed renormalization scheme.

Now let us expand Eq.~(\ref{solD}) in powers of $g$ up to terms
of order $g^2$. We have
\begin{equation}
\label{Dg2}
{\cal D}(p)\approx {\cal D}^{(2)}(p)\equiv
i\frac{\sla{p}+m}{p^2-m^2}-
i\frac{\sla{p}+m}{p^2-m^2}\left[(Z_2-1)(\sla{p}-m)+\delta
m-{\cal M}^{(2)}(p)\right] \frac{\sla{p}+m}{p^2-m^2}
\end{equation}
with
\begin{equation}
\label{sgm1}
{\cal M}^{(2)}(p)=\Sigma(p)+\Sigma_{fc}(p).
\end{equation} 
Applying the CLFD graph technique rules~\cite{cdkm} to the
diagrams shown in Fig.~\ref{fig0a}, we get
$$
\Sigma(p)=-\frac{g^2}{(2\pi)^3}\int (\sla{q}+m)\theta(\omega\cd
q)\delta(q^2-m^2)d^4q\;\theta(\omega\cd k)\delta(k^2-\mu^2)d^4k\, 
\delta^{(4)}(p+\omega\tau-k-q)\frac{d\tau}{\tau-i0} 
$$
\begin{equation}
\label{eq1sen}
= -\frac{g^2}{(2\pi)^3}\int
\frac{(\sla{p}-\sla{k}+\sla{\omega}\tau+m)\;\theta[\omega\cd
(p-k)]}{2\omega\cd (p-k)\tau}\;\frac{d^3k}{2\varepsilon_k}
\equiv g^2\left[{\cal A}(p^2)+{\cal B}(p^2)\frac{\sla{p}}{m}+
{\cal C}(p^2)\frac{m\sla{\omega}}{\omega\cd p}\right]
\end{equation}
and
\begin{equation}\label{sigfc}
\Sigma_{fc}(p)=\frac{g^2\sla{\omega}}
{(2\pi)^3}\int \frac{1}{2\omega\cd
(p-k)}\;\frac{d^3k}{2\varepsilon_k}\equiv
g^2C_{fc}\frac{m\sla{\omega}}{\omega\cd p},
\end{equation}
where
\begin{equation}\label{tau}
\tau=\frac{m^2-(p-k)^2}{2\omega\cd (p-k)}
\end{equation}
and $\varepsilon_k=\sqrt{{\bf k}^2+\mu^2}$.
The functions ${\cal A}(p^2)$, ${\cal B}(p^2)$, ${\cal C}(p^2)$, and
the constant $C_{fc}$ are given in Appendix~\ref{scbos}.
Note that the self-energy $\Sigma(p)$ depends on
the position of the light front~\cite{dugne}, since it contains the
contribution proportional to $\sla{\omega}$. The same contribution was
found earlier in Ref.~\cite{must91}. 
In Eq.~(B9) from Ref.~\cite{must91} it looks as
$\frac{m\gamma^+}{p^+}$, that is just the value of
$\frac{m\sla{\omega}}{\omega\cd p}$ for $\omega=(1,0,0,-1)$.
Substituting the decompositions of $\Sigma(p)$ and $\Sigma_{fc}(p)$
in terms of ${\cal A}(p^2)$, ${\cal B}(p^2)$, ${\cal C}(p^2)$,
and $C_{fc}$, as defined by the r.~h.~s. of Eqs.~(\ref{eq1sen}) 
and~(\ref{sigfc}), into Eq.~(\ref{sgm1}) we can write
\begin{equation}
\label{massopabc}
{\cal M}^{(2)}(p)=g^2\left\{{\cal A}(p^2)+{\cal B}(p^2)\frac{\sla{p}}{m}+
\left[{\cal C}(p^2)+C_{fc}\right]\frac{m\sla{\omega}}{\omega\cd p}\right\}.
\end{equation}

The fermion-boson scattering amplitude calculated in the two-body
approximation discussed here can be obtained by sandwiching
the 2PGF~(\ref{solD}) between the bispinors $\bar{u}(p_1')$ and
$u(p_1)$. Up to terms of order $g^4$ inclusive it is
\begin{equation}
\label{amplg4}
M^{(4)}=ig^2\bar{u}(p_1'){\cal D}^{(2)}(p)u(p_1).
\end{equation}
Substituting Eq.~(\ref{massopabc}) into Eq.~(\ref{Dg2}) and
then into Eq.~(\ref{amplg4}),  
we find after some algebra:
\begin{equation}
\label{amplg41}
M^{(4)}=M^{(4)}_0+M^{(4)}_{\omega},
\end{equation}
where
\begin{eqnarray}
M^{(4)}_0 & = & g^2\bar{u}(p_1')\left\{
\frac{
2m(\sla{p}+m)(\delta m-
g^2[{\cal A}(p^2)+{\cal B}(p^2)+{\cal C}(p^2)+C_{fc}])}
{(p^2-m^2)^2}\right.\nonumber \\
&&\left.
+\frac{[Z_2-2-g^2{\cal B}(p^2)/m](\sla{p}+m)+
\delta m-g^2[{\cal A}(p^2)+{\cal B}(p^2)]}
{p^2-m^2}
\right\}u(p_1), \label{amplm0}
\end{eqnarray}
\begin{equation}
\label{amplmom}
M^{(4)}_{\omega}=
g^4\frac{{\cal C}(p^2)+C_{fc}}{p^2-m^2}\bar{u}(p_1')
\frac{m\sla{\omega}}{\omega\cd p}u(p_1).
\end{equation}
The constants $Z_2$ and $\delta m$ are found in a
standard way (we will specify the corresponding procedure below) 
and do not include any dependence on the orientation of the
light front $\omega$. As a result, the quantity
$M^{(4)}_0$ is also $\omega$-independent.
On the contrary, $M^{(4)}_{\omega}$ explicitly depends on $\omega$
through the term $\bar{u}(p_1')\sla{\omega}u(p_1)$. So, the whole
amplitude $M^{(4)}$ depends on $\omega$ as well, that evidently
calls for a certain additional counterterm (except for the previously
introduced $Z_2$ and $\delta m$) to eliminate this nonphysical
$\omega$-dependence. 

On can also come to the same conclusion from the requirement that the 2PGF~(\ref{Dg2}) 
has a simple pole at $\sla{p}=m$ with the residue equal to $i$. In other words,
there must be
\begin{equation}  
\label{nearpole}
\left.{\cal D}^{(2)}(p)\right|_{\sla{p}\to m}=
\frac{i}{\sla{p}-m}+\ldots=i\frac{\sla{p}+m}{p^2-m^2}+\ldots,
\end{equation}
where the dots designate finite terms. Comparing Eqs.~(\ref{nearpole})
and~(\ref{Dg2}), we have to demand that 
\begin{equation}
\label{eq12bb} 
\lim_{p^2\to m^2}\left\{\frac{(\sla{p}+m)[
(Z_2-1)(\sla{p}-m)+\delta m-{\cal M}^{(2)}(p)](\sla{p}+m)}{p^2-m^2} \right\}=0.
\end{equation}
The expression in the curled brackets in Eq.~(\ref{eq12bb}) is a matrix.
The condition~(\ref{eq12bb}) is imposed on any of  its
matrix element. 
Substituting Eq.~(\ref{massopabc}) into Eq.~(\ref{eq12bb}) and
using the Taylor expansions ${\cal A}(p^2)\equiv A+A'\cdot(p^2-m^2)+\ldots$,
${\cal B}(p^2)\equiv B+B'\cdot(p^2-m^2)+\ldots$, 
${\cal C}(p^2)\equiv C+C'\cdot(p^2-m^2)+\ldots$ near the point $p^2=m^2$,
we can rewrite Eq.~(\ref{eq12bb}) in the form
\begin{eqnarray}
\label{syst1}
&& \left. 2m(\sla{p}+m)\frac{\delta m-g^2(A+B+C+C_{fc})}{p^2-m^2}
\right|_{p^2\to m^2}\nonumber \\
&& + [m(Z_2-1)-g^2B-2g^2m^2(A'+B'+C')+\delta m-g^2(A+B)]
\nonumber\\
&&+ [m(Z_2-1)-g^2B-2g^2m^2(A'+B'+C')]\frac{\sla{p}}{m} +
g^2(C+C_{fc})\frac{\sla{\omega}}{\omega\cd p}=0.
\end{eqnarray}
The absense of the pole term in Eq.~(\ref{syst1}) leads to
\begin{equation}
\label{dmpert}
\delta m=g^2(A+B+C+C_{fc})=
\left.\frac{\bar{u}(p){\cal M}^{(2)}(p)u(p)}{2m}\right|_{p^2=m^2},
\end{equation}
that coincides with the standard perturbative expression. The last
equality immediately follows from Eq.~(\ref{massopabc}) and 
the identities $\bar{u}(p)u(p)=2m$ and $\bar{u}(p)\gamma^{\mu}u(p)=
p^{\mu}\bar{u}(p)u(p)/m$ valid at $p^2=m^2$.
However, it is impossible to choose $Z_2$ in Eq.~(\ref{syst1}) so that
all the other coefficients at the independent matrices $1$, $\sla{p}$,
and $\sla{\omega}$ would be zero, unless $C+C_{fc}=0$. As is seen from
Eqs.~(\ref{Ac}) and~(\ref{Afc}) in
Appendix~\ref{scbos}, the sum $C+C_{fc}$, generally speaking, differs
from zero. Hence,
Eq.~(\ref{syst1}) can not be satisfied. This again indicates the need
for another counterterm in the Lagrangian~(\ref{eq3bba}).

It is easy to see that the counterterm required must have the following
structure (in momentum representation):
\begin{equation}\label{Zom}
Z_{\omega}\left(\frac{m\sla{\omega}}{\omega\cd p}-1\right),
\end{equation}
where $Z_{\omega}$ is a constant to be determined. 
We have supplied this constant with the subscript "$\omega$" in order
to point out its origin, but it does not in fact depend on $\omega$. 
The precise form of Eq.~(\ref{Zom}) is chosen so that this 
counterterm does not contribute when sandwiched between two 
bispinors of equal momenta.

After introducing the new counterterm, the quantity 
$\delta m-{\cal M}^{(2)}(p)$ in 
Eqs.~(\ref{Dg2}) and~(\ref{eq12bb}) must be replaced by
$$
\delta m-{\cal M}^{(2)}(p)+Z_{\omega}\left(
\frac{m\sla{\omega}}{\omega\cd p}-1\right),
$$
that is equivalent to the following two substitutions:
\begin{equation}
\label{subst2}
\delta m\to \delta m-Z_{\omega},\,\,\,\,
g^2[{\cal C}(p^2)+C_{fc}]\to g^2[{\cal C}(p^2)+C_{fc}]-Z_{\omega},
\end{equation}
made in all equations obtained above in this section. Now, 
instead of Eq.~(\ref{syst1}) we get
\begin{eqnarray}
\label{syst1om}
&& \left. 2m(\sla{p}+m)\frac{\delta m-g^2(A+B+C+C_{fc})}{p^2-m^2}
\right|_{p^2\to m^2}\nonumber \\
&& + [m(Z_2-1)-g^2B-2g^2m^2(A'+B'+C')+\delta m-Z_{\omega}-g^2(A+B)]
\nonumber\\
&&+ [m(Z_2-1)-g^2B-2g^2m^2(A'+B'+C')]\frac{\sla{p}}{m} +
[g^2(C+C_{fc})-Z_{\omega}]\frac{\sla{\omega}}{\omega\cd p}=0.
\end{eqnarray}
From here we reproduce for $\delta m$ the same formula~(\ref{dmpert}), 
while for $Z_2$ and $Z_{\omega}$ we find
\begin{equation}
\label{ct2}
Z_2=1+\frac{g^2}{m}B+2mg^2(A'+B'+C')=
\left.1+\frac{1}{2m}
\left\{\bar{u}(p)\frac{\partial{\cal M}^{(2)}(p)}{\partial
\sla{p}}u(p)\right\}\right|_{p^2=m^2},
\end{equation}
\begin{equation}
\label{Zomccfc}
Z_{\omega}=g^2(C+C_{fc}).
\end{equation}
In differentiating, ${\cal M}^{(2)}(p)$ is implied to 
depend
on $\sla{p}$ both explicitly and through $p^2=\sla{p}\cdot \sla{p}$.
Note also that $Z_2$ has the same form as in the standard
renormalization technique.
So, after introducing the additional counterterm~(\ref{Zom}) the behaviour
of the fermion propagator near its pole is indeed governed by 
Eq.~(\ref{nearpole}).

One can easily check that the counterterm~(\ref{Zom}) eliminates the 
$\omega$-dependence of the scattering 
amplitude~(\ref{amplg41})--(\ref{amplmom}).
This nonphysical $\omega$-dependence is concentrated in the term
$M^{(4)}_{\omega}$ given by Eq.~(\ref{amplmom}). The 
substitutions~(\ref{subst2}) with $Z_{\omega}$ taken from Eq.~(\ref{Zomccfc}) 
transform Eq.~(\ref{amplmom}) into
\begin{equation}
\label{amplmom2}
M^{(4)}_{\omega}=\frac{g^4[{\cal C}(p^2)+C_{fc}]-g^2Z_{\omega}}{p^2-m^2}
\bar{u}(p_1')\frac{m\sla{\omega}}{\omega\cd p}u(p_1)=
g^4\frac{{\cal C}(p^2)-{\cal C}(m^2)}{p^2-m^2}
\bar{u}(p_1')\frac{m\sla{\omega}}{\omega\cd p}
u(p_1),
\end{equation}
since, by definition, $C={\cal C}(m^2)$.
As proved in Appendix~\ref{scbos}, ${\cal C}(p^2)$ 
does not depend on $p^2$ [in contrast to the functions
${\cal A}(p^2)$ and ${\cal B}(p^2)$]. 
Therefore $M^{(4)}_{\omega}=0$ and the whole scattering 
amplitude~(\ref{amplg41}) turns out to be $\omega$-independent, as it should.

We have constructed the additional counterterm~(\ref{Zom}) from
perturbative considerations. However, as will be shown below,
it retains its form for the nonperturbative problem discussed in the
present paper. Moreover, its structure is the same for both the scalar
and gauge boson cases. 
In the next section and in Appendix~\ref{lfh} we will explain
how to introduce it in the light-front Hamiltonian.


\section{Scalar boson}\label{scalar}


In this section we give a solution for the state vector of a system
composed of a spin 1/2 fermion 
coupled to a scalar boson, in the Fock decomposition restricted to the
$|F\rangle+|Fb\rangle$ approximation.
A general derivation of the
CLFD fermion-boson Hamiltonian for the scalar boson case 
is presented in Appendix~\ref{scalbos}.

\subsection{Contact terms}\label{contact}

As already mentioned in Sec.~\ref{general}, a peculiarity of LFD is the 
appearence, for particles
with spin, of the so-called  contact interactions~\cite{bpp,cdkm}.
These interactions
arise from the elimination of two non-dynamical degrees of freedom for
the
fermion field, as  recalled in Appendix~\ref{lfh}. In the case of 
spinor-scalar interactions, with the initial interaction Lagrangian
of the form $L^{int}(x)=g\bar{\psi}\psi\varphi$,
where $\psi$ and $\varphi$ are, respectively, the free fermion and
boson fields,
this contact interaction generates a $FFbb\propto g^2$ vertex  in
addition to the usual $FFb$ vertex. In the presence of counterterms
needed to renormalize the theory, like the standard mass counterterm
$\propto \delta m \bar{\psi}\psi$, the elimination of the
non-dynamical degrees of freedom generates in addition two new
contributions to the light-front Hamiltonian.
The first one proportional to  $\delta m^2$
originates from
two self-interaction vertices \footnote{The notation $\delta m^2$ means
$(\delta m)^2$ and
not $\delta(m^2)$}, the second one proportional to $g\delta m$
appears from the product of the self-interaction times the ordinary $FFb$
vertex. Taking into account the specific counterterm discussed in 
Sec.~\ref{need} and introducing, for convenience, a shifted mass 
counterterm $\delta m'$ defined by 
\begin{equation}
\label{dmprime}
\delta{m'} = \delta{m} - Z_{\omega},
\end{equation}
we obtain the final form of the light-front interaction Hamiltonian.
The details of the derivation are given in Appendix~\ref{scalbos}.
We thus get
\begin{equation}
\label{hamiltonian}
H^{int}(x)=H_1(x)+H_{1c}(x)+H_{2}(x)+H_{2c}(x)+H^\prime_{2c}(x)+H_{3}(x)
\end{equation}
with
\begin{subequations}
\begin{eqnarray}
H_1(x)&=&-g\bar{\psi}\psi\varphi,\label{h1} \\
H_{1c}(x)&=& g^2\bar{\psi}\varphi\frac{\sla{\omega}}
{2i(\omega\cd\partial)}\varphi\psi,\label{h1c} \\
H_{2}(x)&=&-\delta m' \bar{\psi}\psi,\label{h2c} \\
H_{2c}(x)&=&g \delta m' \bar{\psi}\left[
\varphi\frac{\sla{\omega}}{2i(\omega\cd\partial)}
+\frac{\sla{\omega}}{2i(\omega\cd\partial)}\varphi\right]\psi,\label{h3c} \\
H^\prime_{2c}(x)&=&\delta
{m'}^2\bar{\psi}\frac{\sla{\omega}}{2i(\omega\cd\partial)}
\psi,\label{h4c} \\
H_{3}(x)&=&-Z_\omega \bar{\psi}\frac{m \sla{\omega}}{i(\omega\cd\partial)}
\psi.\label{h5c}
\end{eqnarray}
\end{subequations}
The Hermitian operator $\frac{1}{i(\omega\cd\partial)}$, which acts on any
coordinate space function $f(x)$ standing to the right of it, turns, in
momentum space, to the factor $\frac{1}{\omega\cd k}$, where $k$ is the
momentum conjugated to the coordinate $x$. The formal definition of
$\frac{1}{i(\omega\cd\partial)}$ in coordinate space is given in
Appendix~\ref{scalbos}. The physical mass of the constituent fermion
(boson) is $m$ ($\mu$). 
The terms $H_1(x)$ and $H_{1c}(x)$ in
Eq.~(\ref{hamiltonian}) are the tree level interactions: the usual
elementary $FFb$ fermion-boson interaction and the contact term on the
fermion line between two elementary interaction vertices. The terms $H_{2}(x)$,
$H_{2c}(x)$, and $H^\prime_{2c}(x)$ are associated with the mass
counterterm we mentioned above.

The last, $\omega$-dependent, term $H_{3}(x)$
plays a particular role. Actually, we postulate its form
on the basis of the observations made in Sec.~\ref{need}.
It is needed to eliminate nonphysical dependence of calculated
observables on the light front position and may 
come both from the
Fock space truncation and from cutting off divergent integrals.
A strict derivation of $H_{3}(x)$ from the "first
principles" is however beyond the scope of the present article and should be a
subject of another work.

We emphasize that $\delta m$ (but not $\delta m'$ !)
is the difference between the dressed and the
bare masses [see Eq.~(\ref{mm0}) above].
Therefore, it represents the shift of mass 
(from the bare to the dressed one) due to
interaction. Since, in gauge theory, 
the corresponding equations of motion are
gauge invariant, the value of $\delta m$ defined in this way is gauge
invariant too, though, generally speaking, the counterterms are not
obligatory gauge invariant. The gauge invariance of $\delta m$ will be
analyzed in our calculations.

We shall apply the Hamiltonian~(\ref{hamiltonian}) in
order to find the composite fermion state vector in the two-body
approximation. For this purpose, we must know the matrix elements of
$H^{int}(x)$ between $|F\rangle$ and $|Fb\rangle$ states. 
In practice, it is convenient to represent these matrix elements 
graphically by means of CLFD diagrams, using
for calculation of their amplitudes the graph technique rules~\cite{cdkm}. 
According to the latters,
matrix elements of an interaction Hamiltonian
are given by amplitudes of the corresponding diagrams taken with the opposite sign. 
In the following we will refer to these amplitudes as interaction vertices.

The interaction vertices for the Hamiltonian~(\ref{hamiltonian}) 
are shown graphically in {Figs.~{\ref{fig1}--\ref{fig6}}}
\footnote{Note that two additional contributions should be added in
Fig.~\ref{fig2} if higher Fock states are considered. They correspond to
the case where the both boson lines originate either from the initial or
final state.}. In each diagram the four-vector $p$ is the total
four-momentum of the initial state. As previously, the
solid and wavy lines correspond to
fermion and boson states, respectively. The dashed lines correspond to 
spurions --- fictitious particles ensuring four-momentum conservation at
each vertex~\cite{cdkm}. There exist three different "elementary"
vertices: a three-point vertex proportional to $g$ and two two-point ones
proportional to $\delta m$ and $Z_{\omega}$. 
We will denote these vertices by a
small full blob, a cross and a cross in an open circle, respectively. The
contact interaction between two elementary vertices, proportional to
$-\frac{\sla{\omega}}{2(\omega\cd k)}$ in momentum space, is denoted by a
fermion line with a large full blob (here $k$ is the four-momentum
corresponding to the line going through the large full blob).

The interaction vertices of $H_1(x)$,  Eq.~(\ref{h1}), are shown in
Fig.~\ref{fig1}.
They describe transitions $|F\rangle\to |Fb\rangle$ and $|Fb\rangle\to
|F\rangle$. The term $H_{1c}(x)$, Eq.~(\ref{h1c}),
in our truncated Fock space, conserves the
number of particles. It gives rise to transitions $|F\rangle\to
|F\rangle$
and $|Fb\rangle\to |Fb\rangle$ only. The corresponding interaction vertices
are shown in Fig.~\ref{fig2}. Note that the amplitude of
the diagram~(c)
is infinite since it is proportional to a divergent integral over
internal boson momentum. The term $H_{2}(x)$, Eq.~(\ref{h2c}), produces
the only matrix element between one-fermion 
states, as shown in Fig.~\ref{fig3}.
The term $H_{2c}(x)$, Eq.~(\ref{h3c}), has four matrix elements
indicated in Fig.~\ref{fig4},
for transitions $|Fb\rangle\to |F\rangle$ and $|F\rangle\to |Fb\rangle$.
The terms $H^\prime_{2c}(x)$, Eq.~(\ref{h4c}), and $H_{3}(x)$,
Eq.~(\ref{h5c}), do not contain bosonic field operators and produce only
one matrix element each (Figs.~\ref{fig5} and~\ref{fig6}, respectively).

To avoid any misunderstanding, we emphasize that, in agreement with the
rules of the
graph technique for CLFD diagrams~\cite{cdkm}, the lines with large
full blobs
representing contact interaction do not correspond to 
propagators, but should
be considered as some "complex" vertices, being
the products of the factors standing at "elementary" vertices. 
For example, such a line in
Fig.~\ref{fig2}(c) is given by
\begin{equation}
g\left[-\frac{\sla \omega}{2\omega \cd (p-k)}\right]g,
\end{equation}
where $p-k$ is just the momentum going through the blob. This momentum
is completely
defined by the energy-momentum conservation at the vertices, including the
spurion four-momentum.

It is worth mentioning that the spin structures of the matrix elements
corresponding  to the diagrams
shown in Figs.~\ref{fig2}(c) and~\ref{fig6} are exactly the
same. Indeed, these matrix elements (denote them by $\langle H_{1c}\rangle$ 
and
$\langle H_3\rangle$, respectively) are given
by the following analytical expressions:
\begin{equation}
\label{M2c,M6} 
\langle H_{1c}\rangle=
g^2C_{fc}\bar{u}'\frac{m\sla{\omega}}{\omega\cd p}u,
\quad 
\langle H_3\rangle=-Z_{\omega}\bar{u'}\frac{m\sla{\omega}}{\omega\cd p}u,
\end{equation}
where
\begin{equation}
\label{cfc2}
C_{fc}=\frac{1}{16\pi^3m}\int
\frac{1}{1-\frac{\omega\cd k}{\omega\cd p}}\,
\frac{d^3k}{2\varepsilon_k},
\end{equation}
as follows from Eq.~(\ref{sigfc}),
and $u$ ($u'$) is the initial (final) fermion bispinor. 
The constant $C_{fc}$ is
calculated in an explicit form in Appendix~\ref{scbos}.
Since $H_{1c}(x)$ and $H_{3}(x)$ come into the full
Hamiltonian~(\ref{hamiltonian}) as a sum, the same property takes place
for their matrix elements shown in Figs.~\ref{fig2}(c) and~\ref{fig6}. If
we define a new constant
\begin{equation}\label{Z}
Z=Z_\omega-g^2C_{fc},
\end{equation}
the total contribution of $\langle H_{1c}\rangle$ and 
$\langle H_3\rangle$ becomes
\begin{equation}
\label{M2c+M6} 
\langle H_{1c}\rangle+\langle H_3\rangle=-Z\
\bar{u'}\frac{m\sla{\omega}}{\omega\cd p}u.
\end{equation}
So, the divergent bosonic loop in Fig.~\ref{fig2}(c) is effectively
"swallowed" by the counterterm proportional to $Z$.


\subsection{Equations in the $|F\rangle+|Fb\rangle$
approximation}\label{eq_scal}

In truncated Fock space we consider in this study
the state vector $\phi_{\sigma}(p)$ includes two sectors only.
Its Fock decomposition has thus the form
\begin{eqnarray}\label{fock}
\phi_{\sigma}(p)&=&\frac{(2\pi)^{3/2}}{\sqrt{N}}\sum_{\sigma'}\int
\phi_{1,\sigma\sigma'}(p_1,p,\omega\tau_1) a^{\dag}_{\sigma'}({\bf
p}_1)|0\rangle
\nonumber\\
&\times&\delta^{(4)}(p_1-p-\omega\tau_1)2(\omega\cd p)
\frac{d\tau_1 d^3p_1}{(2\pi)^{3/2}\sqrt{2\varepsilon_{p_1}}}
\nonumber\\
&+&\frac{(2\pi)^{3/2}}{\sqrt{N}}\sum_{\sigma'}\int \phi_{2,\sigma\sigma'}
(k_1,k_2,p,\omega\tau_2) a^{\dag}_{\sigma'}({\bf k}_1)c^{\dag}({\bf
k}_2)|0\rangle
\nonumber\\
&\times&\delta^{(4)}(k_1+k_2-p-\omega\tau_2)2(\omega\cd p)
\frac{d\tau_2
d^3k_1d^3k_2}{(2\pi)^{3/2}\sqrt{2\varepsilon_{k_1}}(2\pi)^{3/2}
\sqrt{2\varepsilon_{k_2}}},
\end{eqnarray}
where $a^{\dag}$ and $c^{\dag}$ are the creation operators of the free
fermion and boson, respectively, and
$\varepsilon_{p_1}=\sqrt{{\bf p}_1^2+m^2}$,
$\varepsilon_{k_i}=\sqrt{{\bf k}_i^2+m_i^2}$, with the appropriate
masses. The quantities $\phi_1$ and $\phi_2$ are just the light-front
one- and two-body Fock components (wave functions), $N$ is the
normalization factor.

The normalization condition for the state vector has the form
\begin{equation}
\label{norm1}
\phi^{\dag}_{\sigma'}(p')\phi_{\sigma}(p)=(2\pi)^3 \delta_{\sigma\sigma'}2\varepsilon_p
\delta^{(3)}({\bf p}-{\bf p}'),
\end{equation}
that gives for the wave functions~\cite{cdkm}:
\begin{eqnarray}
\label{normwf}
&&\sum_{\tilde{\sigma}}
\phi^{\dag}_{1,\sigma'\tilde{\sigma}}(p_1,p,\omega\tau_1)
\phi_{1,\sigma\tilde{\sigma}}(p_1,p,\omega\tau_1)
+\frac{1}{(2\pi)^3}
\sum_{\tilde{\sigma}}\int\,
\phi^{\dag}_{2,\sigma'\tilde{\sigma}}(k_1,k_2,p,\omega\tau_2)
\phi_{2,\sigma\tilde{\sigma}}(k_1,k_2,p,\omega\tau_2)
\nonumber\\
&&\times\delta^{(4)}(k_1+k_2-p-\omega\tau_2)2(\omega\cd p)d\tau_2
\frac{d^3k_1}{2\varepsilon_{k_1}}\frac{d^3k_2}{2\varepsilon_{k_2}}
=N\delta_{\sigma\sigma'}.
\end{eqnarray}

The state vector corresponds to a total spin of $1/2$ and
satisfies the following equation~\cite{bernard}:
\begin{equation}
\label{shrodeq}
2(\omega\cd
p)\int\tilde{H}^{int}(\omega\tau)\frac{d\tau}{2\pi}\phi_{\sigma}(p)
=-\left[(\hat{P}^0)^2-M^2\right]\phi_{\sigma}(p),
\end{equation}
where $\hat{P}^0$ is the free momentum operator, and
\begin{equation}
\label{hamomt}
\tilde{H}^{int}(\omega\tau)=\int H^{int}(x)e^{-i\tau\omega\cd x}d^4x.
\end{equation}
The interaction Hamiltonian $H^{int}(x)$ is given by
Eq.~(\ref{hamiltonian}). The mass of the composite fermion is denoted by
$M$. In the end of the calculation, we shall take the limit $M \to
m$.

Following Ref.~\cite{bernard}, we define the one- and two-body vertex
functions
$\Gamma_1$ and $\Gamma_2$ by
\begin{subequations}
\begin{eqnarray}
\label{vertex1}
\bar{u}_{\sigma'}(p_1)\Gamma_1u_{\sigma}(p)&=&2(\omega\cd p)\tau_1
\phi_{1,\sigma\sigma'}(p_1,p,\omega\tau_1),\\
\label{vertex2}
\bar{u}_{\sigma'}(k_1)\Gamma_2u_{\sigma}(p)&=&
2(\omega\cd p)\tau_2 \phi_{2,\sigma\sigma'}(k_1,k_2,p,\omega\tau_2).
\end{eqnarray}
\end{subequations}
Due to the delta-functions coming into the decomposition~(\ref{fock}), we
have
$p_1=p+\omega\tau_1$ and $k_1+k_2=p+\omega\tau_2$, that gives
\begin{equation}
\label{tau1tau2}
\tau_1=\frac{m^2-M^2}{2\omega\cd p},\,\,\quad 
\tau_2=\frac{(k_1+k_2)^2-M^2}{2\omega\cd p}.
\end{equation}
Now Eq.~(\ref{shrodeq})
can be rewritten as
\begin{equation}
\label{verteq}
{\cal G}(p)=-\int\tilde{H}^{int}(\omega\tau)\ \hat{\tau}^{-1}{\cal G}(p)\
\frac{d\tau}{2\pi},
\end{equation}
where ${\cal G}(p)$ is defined by a decomposition analogous to
Eq.~(\ref{fock}), with the wave functions replaced by the vertex
functions according to Eqs.~(\ref{vertex1}),~(\ref{vertex2}).
The operator $\hat{\tau}^{-1}$ acting on ${\cal G}(p)$ multiplies
its one-body part by $1/\tau_1$ and its two-body part by $1/\tau_2$.

The system of equations for the vertex functions
$\Gamma_1$ and $\Gamma_2$ is represented graphically  in
Fig.~\ref{fig4a}.
The loop integrals contain $\Gamma_2$ in the integrands. As we shall
check in the end of the calculation, $\Gamma_2$ is a constant for our
Fock space truncation, so that we
can extract it from the integrals. The latters are thus reduced to the
on-mass-shell self-energy 
$\Sigma$ given by Eq.~(\ref{eq1sen}) at $p^2=M^2$. 
Since we should ultimately take the limit $M\to m$, to calculate
$\Sigma$ we may put in Eq.~(\ref{eq1sen}) $p^2=m^2$ at once.
Finally we arrive at the following system of matrix 
equations:
\begin{eqnarray}
\bar{u}(p_1)\Gamma_1 \,u(p)& =&
\bar{u}(p_1)g(-\bar{\Sigma})\,\Gamma_2\, u(p)
   \nonumber\\
   &+& \bar{u}(p_1)\delta m' \frac{\sla{p}+
   \sla{\omega}\tau_1 +m}{m^2-M^2}\Gamma_1 u(p)
   \nonumber\\
   &+&\bar{u}(p_1)\left(\ Z \frac{m\sla{\omega}}{\omega \cd p} \right)
\frac{\sla{p}+\sla{\omega}\tau_1 +m}{m^2-M^2}\Gamma_1 u(p)
   \nonumber\\
& +&   \bar{u}(p_1)g\,\delta m' \left(-\frac{\sla{\omega}}{2\omega\cd p}
\right)
(-\bar{\Sigma})\,\Gamma_2\, u(p) \label{eq1}\nonumber\\
&+&\bar{u}(p_1)\delta {m'}^2 \left(-\frac{\sla{\omega}}{2\omega\cd p}
\right) \frac{\sla{p}+ \sla{\omega}\tau_1 +m}{m^2-M^2}\Gamma_1 u(p),
\end{eqnarray}
\begin{eqnarray}
\bar{u}(k_1)\Gamma_2 u(p)&=&\bar{u}(k_1)g
\frac{\sla{p}+\sla{\omega}\tau_1 +m}{m^2-M^2}\Gamma_1 u(p)
\nonumber\\
&+& \bar{u}(k_1)g^2\left(- \frac{\sla{\omega}}{2\omega\cd p}\right)
(-\bar \Sigma) \,\Gamma_2 u(p) \label{eq2}
\nonumber\\
&+&\bar{u}(k_1)g\,\delta m' \left(- \frac{\sla{\omega}}{2\omega\cd
p}\right) \frac{\sla{p}+\sla{\omega}\tau_1 +m}{m^2-M^2}\Gamma_1 u(p).
\end{eqnarray}
In these equations $p^2=M^2$.
For convenience, we have denoted by $\bar\Sigma$ the fermion self-energy
amputated from the coupling constant:
\begin{equation} \label{sigbar}
\Sigma=g^2 \bar \Sigma.
\end{equation}
Each term in Eqs.~(\ref{eq1}) and~(\ref{eq2}) is in one-to-one
correspondence
with the graphs in Fig.~\ref{fig4a}.
These graphs include various interaction vertices shown in
Figs.~\ref{fig1}--\ref{fig6}, but not all of them. Firstly, we
disregarded the graphs with the vertex shown in Fig.~\ref{fig2}(b), since
they contain a three-body (fermion plus two bosons) intermediate state
which is beyond our two-body approximation. Secondly, by similar reasons,
we neglected the graphs with the vertices shown 
in Figs.~\ref{fig4}(c,d). Indeed, as already
explained in Ref.~\cite{bernard}, the counterterm represented by a line
with a cross
is nothing else than a correction to the fermion self-energy
indicated in Fig.~\ref{fig0a}(a). The self-energy involves the two-body
intermediate state.
Therefore, if we have simultaneously one more boson,
as shown, for instance, in Figs.~\ref{fig4}(c,d), such a graph can be
considered as a
correction to a three-body contribution. Since the latter is out of the
two-body approximation,
we should not take into account the graphs in Figs.~\ref{fig4}(c,d).
In the system of equations~(\ref{eq1}) and~(\ref{eq2}) we also combined the
contributions from graphs
including the interaction vertices shown in
Figs.~\ref{fig2}(c) and~\ref{fig6}
into a single term proportional to $Zm\sla{\omega}/(\omega\cd p)$, 
according to the discussion in the end of Sec.~\ref{contact}. 
For simplicity we kept for this term the same graphical notation
(a cross in an open circle) as for the initial counterterm 
with $Z_{\omega}$.

Using Eq.~(\ref{eq1sen}), 
the fermion on-mass-shell self-energy $\Sigma$ can be represented as
\begin{equation}\label{eq8}
\Sigma = \left.\Sigma(p)\right|_{p^2=m^2}=
g^2A+g^2B\frac{\sla{p}}{m}+g^2C \frac{m\sla{\omega}}{\omega \cd p},
\end{equation}
where, as previously, $A={\cal A}(m^2)$, $B={\cal B}(m^2)$, and
$C={\cal C}(m^2)$.
We display explicitly the coupling constant in the decomposition of the
self-energy. Note that in the general decomposition~(\ref{eq8}) the term
${\sla \omega} {\sla p}$ is also possible, but it does not arise when the
self-energy is given by Eq.~(\ref{eq1sen}), i.e., when it is linear in the
Dirac matrices.  
The coefficients $A$, $B$, and $C$ are calculated
in Appendix~\ref{scbos}.


\subsection{Solution}\label{solut1}

The one-body vertex function $\Gamma_1$ is proportional to a unit matrix.
To avoid the pole terms in Eqs.~(\ref{eq1}) and~(\ref{eq2}), 
it is more appropriate to introduce, instead of $\Gamma_1$,
the quantity $a_1$ defined by
\begin{equation}\label{eqf3}
a_1=\frac{\Gamma_1}{m^2-M^2}
\end{equation}
and related to the one-body wave function by 
$a_1\bar{u}_{\sigma'}(p_1)u_{\sigma}(p)=\phi_{1,\sigma\sigma'}(p_1,p,\omega\tau_1)$.

Substituting $\Gamma_1$ from Eq.~(\ref{eqf3}) into
Eqs.~(\ref{eq1}) and~(\ref{eq2}) and taking the
limit $M\to m$, we get (the fermion spin indices are omitted for shortness)
\begin{eqnarray}
0&=&\bar{u}(p_1)g\left(1-\frac{\sla{\omega}} {2\omega\cd p}\delta m'
\right)(-\bar \Sigma)\,\Gamma_2\, u(p)
\nonumber\\
&+& \bar{u}(p_1)\delta m' \left(1-\frac{\sla{\omega}} {2\omega\cd p}\delta
m' \right)2ma_1 u(p)
\nonumber\\
&+&\bar{u}(p_1)\left( Z \frac{m\sla{\omega}}{\omega \cd p} \right)
2ma_1 u(p),
\label{eq4}\\
\bar{u}(k_1)\Gamma_2 u(p)&=& \bar{u}(k_1)
g\left(1-\frac{\sla{\omega}}{2\omega\cd p}\delta m' \right) 2ma_1 u(p)
\nonumber\\
&+&\bar{u}(k_1)g^2\left(- \frac{\sla{\omega}}{2\omega\cd p}\right)
(-\bar \Sigma)\,\Gamma_2 u(p).
\label{eq5}
\end{eqnarray}
From these homogeneous equations we should find $\delta m', Z$, and (up to
a common factor) $a_1$ and $\Gamma_2$. In contrast to $\Gamma_1$, the
two-body vertex function $\Gamma_2$ has non-trivial spin structure. 
We represent it in terms of scalar components:
\begin{equation}
\bar{u}(k_1)\Gamma_2 u(p)=b_1\bar{u}(k_1)u(p)+
b_2\frac{m}{\omega\cd p}\bar{u}(k_1)\sla{\omega}u(p),
\label{eq7}
\end{equation}
where $b_{1,2}$ are some constants to be determined.
All other possible spin structures are reduced to those listed above. The
counting
rule which gives the total number $N$ of independent scalar components in 
the two-body vertex is
$N=(2s_1+1)(2s_2+1)/2=2\times 2/2=2$.  The structure
$\bar{u}(p_1)\sla{\omega}u(p)$ for the one-body Fock component
is not independent since
$$\bar{u}(p_1)\sla{\omega}u(p)
=\frac{1}{\tau_1}\bar{u}(p_1)\sla{\omega}\tau_1 u(p)
=\frac{1}{\tau_1}\bar{u}(p_1)(\sla{p}_1-\sla{p}) u(p)=
\frac{2\omega\cd p}{m+M}\bar{u}(p_1)u(p).$$

We can easily eliminate the spinors in Eqs.~(\ref{eq4}) and~(\ref{eq5})
by the replacement
$\bar{u}(k) \ldots u(l)\to (\sla{k}+m) \ldots(\sla{l}+m)$, for any
momenta
$k$ and $l$. This amounts to multiply both sides  by $u(k)$ to the left
and by $\bar{u}(l)$ to the right, and then sum over spin projections.
After that we take the trace of each equation. 
In order to get the third equation, we multiply the second
one by $\sla{\omega} m/(\omega\cd p)$ and calculate the trace again.
The results are expressed
through scalar products of the four-vectors entering
these equations. In the limit $M\to m$, the scalar products
are given by
\begin{equation}
\label{scprod1}
p\cd p_1= m^2,\quad
p\cd k_1= m^2-\mu^2/2 +(1-x)(s-m^2)/2,
\end{equation}
with
\begin{equation}
\label{scprod2}
s=(k_1+k_2)^2= (p+\omega\tau_2)^2=m^2+2(\omega\cd p)\tau_2, 
\quad x=\frac{\omega\cd k_1}{\omega\cd p},
\end{equation}
where $k_1$ and $k_2$ are the momenta of the two-body system constituents, 
i.~e. the arguments of
$\Gamma_2=\Gamma_2(k_1,k_2,p,\omega \tau_2)$. Note that because of the
different conservations laws the fermion
momenta $p_1$ (which comes into the one-body Fock component) and $k_1$  
(which comes into
the two-body Fock component) have different scalar products with $p$.

The system of three homogeneous equations for the three
unknown constants $a_1$, $b_1$, and  $b_2$ we thus get is adduced in
Appendix~\ref{eqscb}.
The determinant of this system is 
\begin{multline}
\label{detsc}
Det=-4m^2(m^2-sx)(1-x)\left\{
\delta {m'}^2-2(m+g^2A)\delta m'\right.\\
\left. -g^2[B(g^2B+2g^2C-2Z)-g^2A^2]
+2m(g^2A+g^2B+g^2C-Z)\vphantom{\delta {m'}^2}\right\}.
\end{multline}
In order the system possesses nontrivial solutions, its determinant must be zero.
Equating $Det$ to zero, we
obtain a quadratic equation for $\delta m'$. It has the following
solution:
\begin{equation}\label{eq12}
\delta m' = g^2A +m-  \sqrt{(m-g^2B)\left[m-g^2B- 2(g^2C - Z)\right]}.
\end{equation}
For the other solution (with $+\sqrt{\ldots}$), $\delta m'$ does not tend
to zero when $g\to 0$. We choose therefore the solution~(\ref{eq12}).
Substituting this $\delta m'$ into the system of
equations~(\ref{eq_g10})--(\ref{eq_g2b0}), we can express $b_1$ and $b_2$
through $a_1$:
\begin{equation}\label{eq12a}
b_1=2gm a_1,\quad b_2 =-gm a_1\left(1- \sqrt{1-\frac{2(g^2C -
Z)}{m-g^2B}}\right).
\end{equation}
We see that both $b_1$ and $b_2$ do not depend on
$s$ and $x$. This justifies the extraction of $\Gamma_2$ 
from the integrals
when we derived Eqs.~(\ref{eq4}) and~(\ref{eq5}). The solutions~(\ref{eq12})
and~(\ref{eq12a}) are nonperturbative, since their expansions in
powers of the coupling constant $g$ contain an infinite number of terms.
Eqs.~(\ref{eq4}) and~(\ref{eq5}) are valid for any 
$g$ (for the given Fock space truncation) provided the expressions
under the square roots are non-negative.

According to Eqs.~(\ref{vertex2}) and~(\ref{eq7}), 
the components which determine the 
two-body wave function
are obtained by dividing $b_1$ and $b_2$ by
the factor
$2(\omega\cd p)\tau_2=(s-m^2)$. As a result,
the wave function depends on $\omega$ through the term
$$
\frac{b_2}{s-m^2}\;\frac{m}{\omega\cd p}\bar{u}(k_1)\sla{\omega}u(p).
$$
This dependence must disappear in any physical observable, in
particular, in the residue of the wave function at the pole $s=m^2$. Since,
as follows from the second of Eqs.~(\ref{eq12a}),
$b_2$ is a constant, the only way to ensure this property is to require
$b_2=0$. From here we find the constant $Z=g^2C$ and
\begin{equation}
\label{mpsc}
\delta m' = g^2(A+B).
\end{equation}
Taking into account Eqs.~(\ref{dmprime}) and~(\ref{Z})
yields the final solution
\begin{equation}
\label{eqscb1}
b_1=2gma_1,\quad b_2=0,
\end{equation}
\begin{equation}
\label{eq12b}
\delta m = g^2(A + B+C+C_{fc}), \quad  Z_{\omega}=g^2(C+C_{fc}).
\end{equation}
The three constants $a_1$, $b_{1-2}$ determine the one- and two-body
vertex functions $\Gamma_1$ and $\Gamma_2$. The components $\phi_1$,
$\phi_2$ of the state vector~(\ref{fock}) are related to $\Gamma_1$, 
$\Gamma_2$ 
by Eqs.~(\ref{vertex1}) and~(\ref{vertex2}).
Explicit expressions for $\delta m$ and $Z_{\omega}$ are given in
Appendix~\ref{scbos}.

Substituting the solution for the wave functions into the normalization 
condition~(\ref{normwf}), we find the normalization factor $N$:
\begin{equation}\label{N}
N=4m^2 a_1^2\left\{1+\frac{g^2}{2(2\pi)^3}\int \frac{Tr[(\sla{k}_1+m)
(\sla{p}+m)]}{(s-m^2)^2}
\delta^{(4)}(k_1+k_2-p-\omega\tau_2)2(\omega\cd p)d\tau_2
\frac{d^3k_1}{2\varepsilon_{k_1}}\frac{d^3k_2}{2\varepsilon_{k_2}}
\right\}.
\end{equation}
The integral in Eq.~(\ref{N}) diverges logarithmically. 
Explicit formulas for $N$ are given in Appendix~\ref{scbos}.  
Note that since the factor $\sqrt{N}$ is proportional to $a_1$,
as well as each of the wave functions, the state vector~(\ref{fock})
does not contain the constant $a_1$ at all. So, the final solution for
the state vector is completely determined by the values of
the coupling constant and particle masses.

We remind that we did not renormalize the fermion field
by introducing the factor $Z_2$ as in the first of Eqs.~(\ref{eq2b}).
The fermion field renormalization is not needed in our approach, 
because the factor $N$ does 
the same job for the whole state vector. The two factors $N$ and $Z_2$ 
are tightly related 
to each other. Namely, let us define the constant $Z_2$ as
\begin{equation}\label{ZN}
Z_2=\frac{4m^2a_1^2}{N}.
\end{equation}
It turns out that it is expressed through the self-energy in the standard way:
\begin{equation}\label{Z2p}
Z_2^{-1}-1=-\left.\frac{\partial}{\partial \sla{p}}\Sigma(p)\right|_{\sla{p}=m}.
\end{equation}
Indeed,
using the representation~(\ref{eq1sen}) of the self-energy, 
we get
\begin{equation}\label{dsig}
\left.\frac{\partial}{\partial \sla{p}}\Sigma(p)\right|_{\sla{p}=m}=
g^2\left[2m\frac{\partial {\cal A}(p^2)}{\partial p^2}+
2m\frac{\partial {\cal B}(p^2)}{\partial p^2}+
\frac{1}{m}{\cal B}(p^2)\right]_{p^2=m^2}.
\end{equation}
We used the fact (proved in Appendix~\ref{scbos}) that ${\cal C}(p^2)$
does not depend on $p^2$ and that $\frac{\partial}{\partial \sla{p}}
\frac{\sla{\omega}}{\omega\cd p}=-\frac {\sla{\omega}^2}{(\omega\cd p)^2}=0$.
Substituting the expressions (\ref{abccalA}) and (\ref{abccalB}) for
${\cal A}(p^2)$ and ${\cal B}(p^2)$ into Eq.~(\ref{dsig}),
calculating the corresponding integrals and comparing the result with 
Eq.~(\ref{Ni}) for $N$, we find the relation
\begin{equation}
\label{NdS}
\frac{N}{4m^2a_1^2}=1-\left.
\frac{\partial}{\partial \sla{p}}\Sigma(p)\right|_{\sla{p}=m}.
\end{equation}
From Eqs.~(\ref{ZN}) and~(\ref{NdS}) immediately follows Eq.~(\ref{Z2p}).
Note that the latter is just a standard expression  for the traditional 
counterterm $Z_2$ (see, e.~g., Ref.~\cite{iz}).

In perturbation theory, taking into account that $\Sigma(p)$ is of order $g^2$,
we can transform Eq.~(\ref{Z2p}) to
\begin{equation}\label{Z2pert}
Z_2=1+\left.\frac{\partial}
{\partial \sla{p}}\Sigma(p)\right|_{\sla{p}=m}=
\left.1+\frac{1}{2m}
\left\{\bar{u}(p)\frac{\partial{\cal M}^{(2)}(p)}{\partial
\sla{p}}u(p)\right\}\right|_{p^2=m^2},
\end{equation}
where ${\cal M}^{2}(p)$ is defined by Eq.~(\ref{sgm1}). The last equality
follows from the fact that $\partial \Sigma_{fc}(p)/\partial \sla{p}=0$. 
The r.~h.~s. of Eq.~(\ref{Z2pert}) coincides with that of Eq.~(\ref{ct2})
obtained previously in perturbative renormalization technique.

The solution~(\ref{eq12}), before specifying the value of $Z$, 
is true nonperturbative, 
whereas the final solutions~(\ref{mpsc}) and~(\ref{eq12b}) 
coincide with the perturbative ones. 
It is not a surprise since the self-consistent
calculation in the $|F\rangle+|Fb\rangle$ approximation generates all
contributions to $\delta m$ (or $\delta m'$) of order $g^{2}$, and nothing else. 
A real difference
would appear when higher Fock sectors are taken into account.

The value~(\ref{mpsc}) of $\delta m'$ 
is in agreement with the solution given in Ref.~\cite{glp}.


\section{Gauge boson in the Feynman gauge}\label{vec_feyn}


\subsection {Contact terms}
\label{ctfeyn}

We shall follow here very closely the procedure detailed in the
previous section.
The light-front interaction Hamiltonian for a system of spin 1/2 fermion
coupled to gauge bosons is
discussed in Appendix~\ref{lfh}~2. It has the general form
\begin{eqnarray}
H^{int}(x)&=&-g\bar{\psi}\sla{\varphi}\psi+g^2\bar{\psi}\sla{\varphi}
\frac{\sla{\omega}}{2i(\omega\cd\partial)}\sla{\varphi}\psi
\nonumber\\
&&-\delta m' \bar{\psi}\psi + g \delta m' \bar{\psi}\left[\sla{\varphi}
\frac{\sla{\omega}}{2i(\omega\cd\partial)}+
\frac{\sla{\omega}}{2i(\omega\cd\partial)}\sla{\varphi}\right]\psi+ \delta
{m'}^2\bar{\psi}\frac{\sla{\omega}}{2i(\omega\cd\partial)}\psi\nonumber \\
&&-Z_\omega \bar{\psi}\frac{m \sla{\omega}}
{i(\omega\cd\partial)}\psi\nonumber\\
&&+H'(x).
\label{hamiltonian_f}
\end{eqnarray}
and differs from the Hamiltonian~(\ref{hamiltonian}) by the replacement
$\varphi\to\sla{\varphi}$ and by the additional contribution $H'(x)$ which
represents contact terms associated to the gauge field. 
An explicit 
form of $H'(x)$ depends on the gauge. Generally speaking, $H'(x)$ taken
in an arbitrary gauge and expanded
in powers of $g$ gives rise to an infinite number of contact terms, even in
truncated Fock space.

For the Feynman gauge, $H'(x)$ acquires the form
\begin{equation}
\label{hamfeyn}
H'_F(x)=H'_{1F}(x)+H'_{2F}(x)+H'_{3F}(x)
\end{equation}
with
\begin{subequations}
\begin{eqnarray}
\label{hamfeyn1}
H'_{1F}(x)&=&g^2\bar{\psi}\sla{\varphi}\frac{\sla{\omega}}{2i(\omega\cd D)}
\sla{\varphi}\psi,\\
\label{hamfeyn2} H'_{2F}(x)&=&g\delta m'\bar{\psi}\left[
\sla{\varphi}\frac{\sla{\omega}}{2i(\omega\cd D)}+
\frac{\sla{\omega}}{2i(\omega\cd D)}\sla{\varphi}\right]\psi,\\
\label{hamfeyn3} H'_{3F}(x)&=&\delta
{m'}^2\bar{\psi}\frac{\sla{\omega}}{2i(\omega\cd D)}\psi, 
\end{eqnarray}
\end{subequations}
where the Hermitian operator $\frac{1}{i(\omega\cd D)}$ is defined in
Appendix~\ref{fgauge}. An expansion 
of $H'_F(x)$ in powers of $g$ starts with a term of order $g^3$ (we took
into account that the corresponding expansions of 
$\delta m'$ and of the operator $\frac{1}{i(\omega\cd D)}$ begin with terms
of order $g^2$ and $g$, respectively). In full (non-truncated) Fock space
this series does not break and the number of contact terms is infinite.
However, it becomes finite in truncated Fock space. Since we are
working within one- and two-body sectors, we have to know the non-zero
matrix elements of $H_{F}'(x)$ between these sectors. Let us consider
first the term $H'_{1F}(x)$. We can represent $H'_{1F}(x)$ as the following
series in powers of $g$:
\begin{equation}
\label{hamfeynseries}
H'_{1F}(x)=\frac{g^2}{2}\sum_{n_1=0,n_2=0,
n_1+n_2\rangle 0}^{\infty}\frac{(-1)^{n_1}}{n_1!n_2!}g^{n_1+n_2}H_{n_1n_2},
\end{equation}
\begin{equation}
\label{Hn1n2}
H_{n_1n_2}=\bar{\psi}\sla{\varphi}\left(
\frac{1}{i\omega\cd\partial}\omega\cd\varphi\right)^{n_1}
\frac{\sla{\omega}}{i\omega\cd\partial}
\left[\left(\frac{1}{i\omega\cd\partial}\omega\cd\varphi\right)^{n_2}
\sla{\varphi}\psi\right].
\end{equation}
Each term $H_{n_1n_2}$ contains $n_1+n_2+2$ boson field operators
$\varphi_{\nu}$.
When we calculate matrix elements between
states with the fixed number of particles, these operators must be
contracted
with external boson operators or among themselves. It is easy to see that
contractions between two different operators $\omega\cd\varphi$ or
between $\omega\cd\varphi$ and $\sla{\varphi}$ give zero in the Feynman
gauge.
Indeed, since $\langle 0|[\varphi_{\mu},\varphi_{\nu}]|0\rangle
\sim -g_{\mu\nu}$, the "internal" contractions
between the operators $\omega \cd \varphi$ in Eq.~(\ref{Hn1n2}) are
proportional to $\omega^{\mu}\omega^{\nu}g_{\mu\nu}$, that is zero.
Contractions between $\omega \cd \varphi$ and
$\sla{\varphi}$ also disappear because they are proportional to
$\gamma^{\mu}\sla{\omega}\omega^{\nu}g_{\mu\nu}=
\sla{\omega}\sla{\omega}=\omega^2=0$. The only remaining possibility is
to contract the
operators $\omega\cd \varphi$ with the external boson field operators.
The number
of the latters is fixed and does not exceed two in our approximation.
Hence, we must have $n_1+n_2\leq 2$. This just means a
truncation of the
series~(\ref{hamfeynseries}) to a polynomial of the fourth order in $g$.
Evidently, the same arguments hold true for
each of the terms $H'_{2F}(x)$ and $H'_{3F}(x)$, and, hence, for
the operator $H'_F(x)$ as a whole.
Note that
$H'_F(x)$ does not have non-zero matrix elements within the one-body
sector,
because it contains at least one operator $\omega\cd\varphi$.

Considering matrix elements of the operator $H'_F(x)$
between the states $|F\rangle$ and $|Fb\rangle$, we find
that each piece $H'_{1F}(x)$, $H'_{2F}(x)$, or $H'_{3F}(x)$ produces
eight different interaction vertices.
For example, in Fig.~\ref{figmelf} all possible interaction vertices
generated by the piece $H'_{1F}(x)$ are shown.
Each "elementary" vertex (small black blob) corresponds to the factor
$g\gamma^{\mu}$, the propagator of boson with four-momentum $q$ being
$-g_{\mu\nu}\delta(q^2)\theta(\omega\cd q)$. The other notations are the
same as those adopted for the scalar boson case, except for new factors
indicated by large open circles with two fermion
and one boson legs. These factors proceed from
different powers of the operators
$\frac{1}{i\omega\cd\partial}\omega\cd\varphi$ in Eq.~(\ref{Hn1n2}).
A large open circle corresponds to the factor $\pm g \frac{\omega^\mu}{
\omega\cd k}$, where $k$ is the four-momentum of the boson line
connected with this open circle. The sign plus (minus) is taken for
an outgoing (incoming) boson line. Additionally, each matrix element
must be multiplied by a combinatorial factor $1/(n_{1}!n_{2}!)$, where
$n_{1}$  ($n_{2}$) is the number of large open circles standing to the
left (to the right) from the large full blob. We have omitted
two diagrams which differ from those shown in Figs.~\ref{figmelf}(g)
and~(h) by the interchange of the initial and
final boson lines because such an interchange does not affect analytical
expressions
for the corresponding amplitudes. So, to incorporate the contributions
from the
omitted diagrams it is enough to multiply the amplitudes of the diagrams in
Figs.~\ref{figmelf}(g),~(h) by a factor of two, just cancelling the
combinatorial factor $1/2!$

As far as the interaction vertices produced by the terms $H'_{2F}(x)$ and
$H'_{3F}(x)$ are concerned, they can be easily obtained from the diagrams
of
Fig.~\ref{figmelf}. Indeed, to get the vertices for $H'_{2F}(x)$ one
should substitute in all these diagrams one of small full blobs by a cross
and transform the boson loop to a free boson line incoming to
(or outgoing from) the remaining small full blob. In such a way
each of the diagrams in Figs.~\ref{figmelf}(a)--(d) transforms into two
new ones, while the diagrams shown in Figs.~\ref{figmelf}(e)--(h)
give nothing because all of them contain after transformation three
boson legs. The interaction vertices for $H'_{3F}(x)$ are obtained from those
in Fig.~\ref{figmelf} simply by substituting all small full blobs by
crosses and by removing the boson loops.

Note that similarly to the scalar boson case, any line with a large (full
or open) blob, or
with two of them, is not a propagator, but only a product of the factors
corresponding to the elementary vertices (large and small full blobs,
and large open circles). For example, the internal solid line in
Fig.~\ref{figmelf}(a) corresponds to the analytical expression
\begin{equation} \label{prec}
g\gamma^\nu \left[-\frac{\sla \omega}{2\omega \cd (p-q)} \right] \left
( -g \frac{\omega^{\rho}}{\omega \cd k_1}\right) g\gamma^\mu,
\end{equation}
where $p=p_1+k_1-\omega\tau_1=p_1'-\omega\tau_1'$ is the total momentum 
of the initial or final state.
To calculate the full matrix element, one should sandwich Eq.~(\ref{prec}) with
bispinors, then multiply it by the boson propagator
  $-g_{\mu\nu}\delta(q^2)\theta(\omega \cd q)$ and by the polarization
vector
  $e^{\lambda}_{\rho}(k_{1})$, integrate over $d^4q/(2\pi)^3$ and multiply
  the result by $(-1)$.

It is important to note that although we have calculated matrix elements
of $H'_F(x)$ between the states $|F\rangle$ and $|Fb\rangle$, any of the
interaction vertices obtained from $H'_F(x)$ includes intermediate
states with
at least two  bosons or with one boson and a cross. Repeating the same
arguments as those exposed for the scalar boson case, we have thus to conclude
that the whole contribution from $H'_F(x)$ is beyond our two-body
approximation.

\subsection{Equations in the $|F\rangle+|Fb\rangle$
approximation}\label{eq_feyn}

The Fock decomposition of the state vector is similar to the one for
scalar bosons, given by Eq.~(\ref{fock}).  The only difference is the
dependence of the boson creation operator and the Fock components on the
boson spin index $\lambda$.
As usual, we define the one- and two-body vertex functions
$\Gamma_1$ and $\Gamma_2^\mu$ by 
[cf. with Eqs.~(\ref{vertex1}) and~(\ref{vertex2})]
\begin{subequations}
\begin{eqnarray}
\label{vertexg1}
\bar{u}_{\sigma'}(p_1)\Gamma_1u_{\sigma}(p)&=&2(\omega\cd p)\tau_1
\phi_{1,\sigma\sigma'}(p_1,p,\omega\tau_1),\\
\label{vertexg2}
\bar{u}_{\sigma'}(k_1)\Gamma_2^\mu e^\lambda_\mu(k_2) u_{\sigma}(p)&=&
2(\omega\cd p)\tau_2
\phi_{2,\sigma\sigma'}^\lambda(k_1,k_2,p,\omega\tau_2).
\end{eqnarray}
\end{subequations}

We should now establish, in the two-body approximation, the whole set of
matrix
elements of the Hamiltonian~(\ref{hamiltonian_f}) between the one-
and
two-body sectors. Since the Hamiltonian~(\ref{hamiltonian_f}),
excepting for the term $H'_F(x)$, looks quite similar to that
for the scalar boson case, Eq.~(\ref{hamiltonian}), we can make
use of this analogy. 
So, we can take directly the graphs shown in
Figs.~\ref{fig1}--\ref{fig3},~\ref{fig4}(a,b),~\ref{fig5},~\ref{fig6},
substituting $g\to g\gamma^{\mu}$ everywhere, and add to this set 
the contributions from $H'_F(x)$,
indicated in Fig.~\ref{figmelf}, which are compatible with our Fock
space truncation.
However, as was noticed in the end of Sec.~\ref{ctfeyn}, these
contributions
correspond in fact to many-body sectors and
should therefore be disgarded.

For further calculations we will need the fermion self-energy. 
In the Feynman gauge it is
\begin{eqnarray}\label{senf}
\Sigma^F &=& -\frac{g^2}{(2\pi)^3}\int
\frac{(-g_{\mu\nu})\gamma^\mu(\sla{p}-\sla{k}+\sla{\omega}\tau+m)\gamma^\nu
\;\theta[\omega\cd (p-k)]}{2\omega\cd
(p-k)\tau}\;\frac{d^3k}{2\varepsilon_k} \nonumber\\
 &\equiv&
\gamma^\mu\left(\Sigma_{\mu\nu}^F\right)\gamma^\nu\ =
g^2A+g^2B\frac{\sla{p}}{m} +g^2C\frac{m\sla{\omega}}{\omega\cd p}
\end{eqnarray}
with $\tau$ defined by Eq.~(\ref{tau}). Note that the
constants $A$, $B$, and $C$ here do not coincide with those 
for the scalar boson case, defined by Eq.~(\ref{eq8}).
From Eq.~(\ref{senf}) follows
\begin{equation}
\label{selfenfeyn}
\Sigma_{\mu\nu}^F=-g_{\mu\nu}(g^2A_1+g^2B_1\frac{\sla{p}}{m}
+g^2C_1\frac{m\sla{\omega}}{\omega\cd p})\equiv g^2\bar \Sigma_{\mu\nu}^F,
\end{equation}
where 
\begin{equation}\label{aa1}
A=-4A_1,\quad
B=2B_1,\quad C=2C_1.
\end{equation}
The constants $A$, $B$, and $C$ are given by Eqs.~(\ref{Af})--(\ref{Cf}) in 
Appendix~\ref{gbosf}. 

We can now write down a system of equations for the vertex functions.
The
graphical representation of these equations is shown in
Fig.~\ref{figFeyn}.
It is very similar to the one for scalar bosons, as shown
in Fig.~\ref{fig4a}.
The
only formal difference consists in the appearance of four-dimensional
indices in boson propagators and vertices.
Generally speaking, the vertices $\Gamma_1$ and $\Gamma_2^{\mu}$ do have
momentum
dependence. Because of our Fock space truncation, they are however
constants. At the
present stage we take this property as an ansatz, and it will be
justified by the final results.

In an analytical form, this system of equations reads
\begin{eqnarray}
\label{eqg1feyn}
\bar{u}(p_1)\Gamma_1u(p)
& = & {\displaystyle \bar{u}(p_1) g\gamma^{\nu}
(-\bar\Sigma_{\nu\rho}^F)\Gamma_2^{\rho}u(p)}
\nonumber\\
                        & + & {\displaystyle \bar{u}(p_1)\delta m'
\frac{(\sla{p}+\sla{\omega}\tau_1+m)}{m^2-M^2}\Gamma_1u(p)}
\nonumber\\
                           & + & {\displaystyle
\bar{u}(p_1)\left(Z\frac{m\sla{\omega}}
       {\omega\cd p}\right)
\frac{(\sla{p}+\sla{\omega}\tau_1+m)}{m^2-M^2}\Gamma_1u(p)}
       \nonumber\\
                             & + & {\displaystyle \bar{u}(p_1)g\delta
m'\left(-\frac{\sla{\omega}}
       {2 \omega\cd p}\right)\gamma^{\nu}
(-\bar\Sigma_{\nu\rho}^F)\Gamma_2^{\rho}u(p)} \nonumber\\
& + & {\displaystyle \bar{u}(p_1)\delta {m'}^2\left(-\frac{\sla{\omega}}
       {2 \omega\cd p}\right)
\frac{(\sla{p}+\sla{\omega}\tau_1+m)}{m^2-M^2}\Gamma_1u(p)},
\end{eqnarray}
\begin{eqnarray}
\label{eqg2feyn}
\bar{u}(k_1)[\Gamma_2^{\mu}e_{\mu}^{\lambda}(k_2)]u(p)
            & = & {\displaystyle
\bar{u}(k_1)g\gamma^{\mu}e_{\mu}^{\lambda}(k_2)
\frac{(\sla{p}+\sla{\omega}\tau_1+m)}{m^2-M^2}\Gamma_1u(p)} \nonumber\\
& + & {\displaystyle \bar{u}(k_1) g^2\gamma^{\mu}
e_{\mu}^{\lambda}(k_2)
      \left(-\frac{\sla{\omega}}{2 \omega\cd p}\right)\gamma^{\nu}
                   (-\bar\Sigma_{\nu\rho}^F)\Gamma_2^{\rho}u(p)}
\nonumber\\
            & + & {\displaystyle \bar{u}(k_1)g\delta m'\gamma^{\mu}
e_{\mu}^{\lambda}(k_2)
                  \left(-\frac{\sla{\omega}} {2 \omega\cd p}\right)
\frac{(\sla{p}+\sla{\omega}\tau_1+m)}{m^2-M^2}\Gamma_1u(p) }.
\end{eqnarray}
As explained in Sec.~\ref{contact}, we absorbed the (divergent) term
proportional to $Z_\omega$ and the contribution from the diagram
shown in Fig.~\ref{fig2}(c) (with
the scalar boson replaced by the gauge one) in the unique counterterm
proportional to $Z$, defined by Eq.~(\ref{Z}),
where $C_{fc}$ is now given by
\begin{equation}\label{sigfcf}
g^2C_{fc}\frac{m\sla{\omega}}{\omega\cd p}
=\frac{g^2}{(2\pi)^3}\int \frac{(-g_{\mu\nu})\gamma^\mu
\sla{\omega}\gamma^\nu}{2\omega\cd (p-k)}\,\frac{d^3k}{2\varepsilon_k}.
\end{equation}
The constant $C_{fc}$ 
is calculated in Appendix~\ref{gbosf} [see Eq.~(\ref{Cfc})].


\subsection{Solution}\label{solut2}
In order to solve Eqs.~(\ref{eqg1feyn}) and~(\ref{eqg2feyn}), it is
convenient
to decompose the vertex functions in invariant amplitudes, similarly to
what was done
in Sec.~\ref{scalar}. The structure of
$\bar{u}(p_1)\Gamma_1u(p)$ coincides with the one for the spinor-scalar
interaction [cf. with Eq.~(\ref{eqf3})]:
\begin{equation}
\label{gamma1feyn}
\frac{\bar{u}(p_1)\Gamma_1 u(p)}{m^2-M^2}=a_1\bar{u}(p_1)u(p),
\end{equation}
where $a_1$ is a constant and the factor $(m^2-M^2)^{-1}$ reflects the
behaviour of $\Gamma_1$
in the limit $M\to m$. The number of independent invariant amplitudes for
the vertex function
$\bar{u}(k_1)[\Gamma_2^{\mu}e_{\mu}^{\lambda}(k_2)]u(p)$ coincides with
that for the
reaction $1/2+0\to 1/2+1$. However, one should take into account that in
the Feynman gauge
the vector boson wave function has {\em four} independent components. So,
the total number
of invariant amplitudes is $(2\times 2\times 4)/2=8$. We choose the
following set of
invariant amplitudes:
\begin{eqnarray}
\bar{u}(k_1)[\Gamma_2^{\mu}e_{\mu}^{\lambda}(k_2)]u(p)&=&
\bar{u}(k_1)\left[b_1\gamma^{\mu}+b_2\frac{m\omega^{\mu}}{\omega\cd p}+
b_3\frac{m\sla{\omega}\gamma^{\mu}}{\omega\cd p}
+b_4\frac{m^2\sla{\omega}\omega^{\mu}}{(\omega\cd p)^2}
\right. \nonumber \\
&&\nonumber \\
\label{gamma2feyn8}
&+& \left. b_5\frac{p^{\mu}}{m}+b_6\frac{k_1^{\mu}}{m}+
b_7\frac{m\sla{\omega}p^{\mu}}{\omega\cd p}
+b_8\frac{m\sla{\omega}k_1^{\mu}}{\omega\cd p}\right]
e_{\mu}^{\lambda}(k_2)u(p).
\end{eqnarray}
One can easily check that in our approximation the structures
proportional
to $p^{\mu}$ and $k_1^{\mu}$
do not contribute at all. Indeed, whatever term from
Eq.~(\ref{gamma2feyn8})
we substitute into
Eq.~(\ref{eqg2feyn}), the result will always include structures
proportional to
$\gamma^{\mu}, \omega^{\mu}, \sla{\omega}\gamma^{\mu}$, or
$\sla{\omega}\omega^{\mu}$ only,
i.~e. the four structures proportional to $b_{5-8}$ must be absent.
Although all the four coefficients $b_{5-8}$ can be
obtained automatically, as a solution of a system of four linear
equations,
with the result $b_5=b_6=b_7=b_8=0$,
we may set  $b_{5-8}=0$
from the very beginning  in order to simplify the
calculations, and reduce Eq.~(\ref{gamma2feyn8}) to
\begin{equation}
\label{gamma2feyn4}
\bar{u}(k_1)[\Gamma_2^{\mu}e_{\mu}^{\lambda}(k_2)]u(p)
=\bar{u}(k_1)\left[b_1\gamma^{\mu}+b_2\frac{m\omega^{\mu}}{\omega\cd p}+
b_3\frac{m\sla{\omega}\gamma^{\mu}}{\omega\cd p}
+b_4\frac{m^2\sla{\omega}\omega^{\mu}}{(\omega\cd p)^2}
\right]e_{\mu}^{\lambda}(k_2)u(p).
\end{equation}
To solve the equations~(\ref{eqg1feyn}) and~(\ref{eqg2feyn}),
we repeat the same steps as done in Sec.~\ref{solut1} for scalar bosons. We
insert Eqs.~(\ref{gamma1feyn}),~(\ref{gamma2feyn4})
into Eqs.~(\ref{eqg1feyn}),~(\ref{eqg2feyn}), take the limit $M\to m$,
and substitute the spinors and polarization vectors according to
$\bar{u}(p_1)\ldots u(p)\to (\sla{p}_1+m)\ldots (\sla{p}+m)$,
$\bar{u}(k_1)\ldots u(p)\to (\sla{k}_1+m)\ldots (\sla{p}+m)$,
$e_{\mu}^{\lambda}(k_2)\to -g_{\mu\mu'}$. After these transformations
we take the trace from Eq.~(\ref{eqg1feyn}). Then
we multiply the transformed Eq.~(\ref{eqg2feyn}) in turn
by $\gamma^{\mu'}$, $m\omega^{\mu'}/(\omega\cd p)$,
$m\gamma^{\mu'}\sla{\omega}/(\omega\cd p)$,
and $p^{\mu'}/m$, and take traces again.
As a result, we get a system of five
homogeneous linear equations for the five coefficients $a_1$, $b_{1-4}$,
which is given in Appendix~\ref{eqsfb}.
For finding nontrivial solutions of this system, we have to
demand its determinant to be zero:
\begin{multline}
\label{determfeyn} Det=4m^4(m^2-sx)x(1-x^2)\left\{
\delta {m'}^2-2(m-4g^2A_1)\delta m'\right.\\
\left. -4g^2[B_1(g^2B_1+2g^2C_1-Z)-4g^2A_1^2]
-2m(4g^2A_1-2g^2B_1-2g^2C_1+Z)\vphantom{\delta {m'}^2}\right\}=0.
\end{multline}
Solving Eq.~(\ref{determfeyn}) with respect to $\delta m'$, we find two solutions.
Evidently, we should choose the solution which 
gives $\delta m'\to 0$ as $g\to 0$. Expressing $A_1$, $B_1$, and $C_1$ 
through the constants $A$, $B$, and $C$ by means of 
Eqs.~(\ref{aa1}), we get
\begin{eqnarray} \label{delmfeyn}
\delta m'&=&-4g^2A_1+m - \sqrt{(m-2g^2B_1)[m-2g^2B_1-2(2g^2C_1-Z)]}
\nonumber\\
&=&g^2A +m-  \sqrt{(m-g^2B)\left[m-g^2B- 2(g^2C - Z)\right]}.
\end{eqnarray}
Substituting Eq.~(\ref{delmfeyn}) into Eqs.~(\ref{eqF1feyn})--(\ref{eq4feyn}),
we obtain
\begin{equation}
\label{b1feyn}
b_1=2gma_1,
\end{equation}
\begin{equation}
\label{b2feyn} b_2=-2b_3=-2gma_1\left(1- \sqrt{1-\frac{2(g^2C -
Z)}{m-g^2B}}\right), \quad b_4=0.
\end{equation}
Like to the scalar boson case, we require the residue of the 
two-body wave function at the pole $s=m^2$ to be $\omega$-independent. 
This leads to
the conditions $b_2=b_3=b_4=0$ which can be satisfied if
$Z=g^2C$.
We thus arrive to the following final 
solution for the vertex functions and the two counterterms:
\begin{equation}
\label{solgamma2feyn}
b_1=2gma_1,\quad b_{2-8}=0,
\end{equation}
\begin{equation}
\label{soldelmfeyn} 
\delta m=g^2(A+B+C+C_{fc}), \quad Z_{\omega}=g^2(C+C_{fc}).
\end{equation}
Eqs.~(\ref{soldelmfeyn}) formally coincide with Eqs.~(\ref{eq12b})
found for the scalar boson case. However the constants 
$A$, $B$, $C$, and $C_{fc}$ theirselves depend on the boson type.
Due to this reason the exact values of the counterterms $\delta m$
and $Z_{\omega}$ here are different from those in Eqs.~(\ref{eq12b}).
They are given in Appendix~\ref{gbosf}.


\section{Gauge boson in the Light-Cone gauge}\label{vec_lc}


\subsection{Contact terms}

The interaction Hamiltonian has the same
form~(\ref{hamiltonian_f}) as in the Feynman gauge, except for the
gauge dependent term $H'(x)$. In the LC gauge where $\omega\cd
\varphi=0$ we have (see Appendix~\ref{lcg})
\begin{equation}
\label{hamlf}
H'_{LC}(x)=g^2(\bar{\psi}\sla{\omega}\psi)\frac{1}{2(i\omega\cd\partial)^2}
(\bar{\psi}\sla{\omega}\psi).
\end{equation}
In contrast to all other gauges, this contribution to the light-front Hamiltonian
generates only one contact term, of order $g^2$, associated to the
propagator
of the gauge field.

As far as the non-zero matrix elements of $H_{LC}'(x)$ are concerned, the
situation here is simpler than in the case of the Feynman gauge. This part
of the Hamiltonian conserves the number of particles in truncated Fock
space which we consider here. It generates two interaction
vertices in the one-body sector, as shown diagrammatically in
Fig.~\ref{fig6a}.

The fermion contact term with the boson loop, shown 
in Fig.~\ref{fig2}(c) (with the scalar boson replaced by the gauge one)
is now 
\begin{equation}\label{eq3b}
g^2C_{fc}\frac{m\sla{\omega}}{\omega\cd p}=\frac{g^2}{(2\pi)^3}\int
\frac{d_{\mu\nu}\gamma^\mu\sla{\omega}\gamma^\nu}{2\omega\cd
(p-k)}\,\frac{d^3k}{2\varepsilon_k},
\end{equation}
where
\begin{equation}\label{d}
d_{\mu\nu}=-g_{\mu\nu}+\frac{k_{\mu}\omega_{\nu}+
k_{\nu}\omega_{\mu}}{\omega\cd k}
\end{equation}
results from the vector boson propagator in the LC gauge.

The sum of the two (divergent) fermionic loops
shown in Fig.~\ref{fig6a}
represents a contact correction, on the boson line, to the self-energy
contribution. It can be written as
\begin{equation}\label{eq3c}
g^2C_{bc}\frac{m\sla{\omega}}{\omega\cd p}\equiv
\frac{g^2}{(2\pi)^3}\int 
\left(\frac{\sla{\omega}\sla{k}\sla{\omega}}{2[\omega\cd (p-k)]^2}-
\frac{\sla{\omega}\sla{k}\sla{\omega}}{2[\omega\cd (p+k)]^2}\right)\,
\frac{d^3k}{2\varepsilon_k}.
\end{equation}
The expressions~(\ref{eq3b}) and~(\ref{eq3c}) 
have the same matrix structure as the one coming from the
$\omega$-dependent counterterm proportional to $Z_\omega$. 
We can therefore again absorb them, as 
in Eq.~(\ref{Z}), into a unique
term which, for simplicity, we also call $Z$:
\begin{equation}\label{Zpp}
Z=Z_\omega-g^2C_{fc}-g^2C_{bc}.
\end{equation}


\subsection{Equations in the $|F\rangle+|Fb\rangle$
approximation}\label{eq_lc}

The contribution~(\ref{eq3c}) 
should be excluded from the
equation for the last Fock component $\Gamma_2$. The reason for this is
that $H_{LC}'(x)$ defined by Eq.~(\ref{hamlf}) describes corrections to
bosonic propagators. These corrections to $\Gamma_2$ can be regarded
as "foot-prints" of contributions from the three-body sector containing
one fermion and two bosons, which is outside of our model space.

The fermion self-energy has now the form
\begin{eqnarray}\label{senflc}
\Sigma^{LC}&=&-\frac{g^2}{(2\pi)^3}\int
\frac{d_{\mu\nu}\gamma^\mu(\sla{p}-\sla{k}+\sla{\omega}\tau+m)\gamma^\nu
\;\theta[\omega\cd
(p-k)]}{2\omega\cd(p-k)\tau}\;\frac{d^3k}{2\varepsilon_k} \nonumber\\
&\equiv& \gamma^\mu\Sigma^{LC}_{\mu\nu}\gamma^\nu = g^2A +
g^2B\frac{\sla{p}}{m}+ g^2C\frac{m\sla{\omega}}{\omega\cd p},
\end{eqnarray}
where $d_{\mu\nu}$ and $\tau$ are given by Eqs.~(\ref{d}) and~(\ref{tau}),
respectively.

The symmetric tensor $d_{\mu\nu}$ is transversal to both  $\omega$ and the
boson momentum $k$. As a consequence, the operator $\Sigma_{\mu\nu}^{LC}$
in Eq.~(\ref{senflc}) is a symmetric tensor transversal to $\omega$.
Besides that, it contains only zeroth and first powers of
$\gamma$-matrices. Due to all these, the tensor $\Sigma_{\mu\nu}^{LC}$ has
the following general structure:
$$
\Sigma_{\mu\nu}^{LC}=
-\left(
g_{\mu\nu}-\frac{p_{\mu}\omega_{\nu}+p_{\nu}\omega_{\mu}}{\omega\cd p}
\right)\left(g^2A_1+g^2B_1\frac{\sla{p}}{m}\right)+
g^2C_1\frac{m}{\omega\cd p}
\left(g_{\mu\nu}\sla{\omega}-\gamma_{\mu}\omega_{\nu}-
\gamma_{\nu}\omega_{\mu}\right)
$$
\begin{equation}\label{selfenlf}
+g^2C_1'\frac{m}{\omega\cd p}\left(
\sla{\omega}\frac{p_{\mu}\omega_{\nu}+p_{\nu}\omega_{\mu}}{\omega\cd p}-
\gamma_{\mu}\omega_{\nu}-\gamma_{\nu}\omega_{\mu}\right)
+\frac{m^2\omega_{\mu}\omega_{\nu}}{(\omega\cd p)^2}
\left(g^2A_2+g^2B_2\frac{\sla{p}}{m}+
g^2C_2\frac{m\sla{\omega}}{\omega\cd p}\right),
\end{equation}
determined by seven scalar functions $A_1$, $B_1$, $C_1$, $C_1'$,
$A_2$, $B_2$, and $C_2$.
As usual, we define $\Sigma_{\mu\nu}^{LC} = g^2 \bar
\Sigma_{\mu\nu}^{LC}$. Multiplying Eq.~(\ref{selfenlf}) by $\gamma^{\mu}$
to the left and by $\gamma^{\nu}$ to the right, contracting
over the indices $\mu$ and $\nu$, and comparing
the result with the r.~h.~s. of Eq.~(\ref{senflc}), we get
\begin{equation}
\label{ctilde} A=-2A_1,\quad B=2B_1,\quad C=2(B_1+B_2-5C_1-4C'_1).
\end{equation}
The values of $A$, $B$, and $C$ are given in Appendix~\ref{gboslc} by 
Eqs.~(\ref{eq5bA})--(\ref{eq5bC}).

We thus arrive, in the LC gauge, at a system of equations for
the vertex functions, which is quite similar to that
for the case of gauge bosons in the Feynman gauge, shown graphically in
Fig.~\ref{figFeyn}.
Because $H'_{LC}(x)$ does not
produce any matrix elements of new structure as compared with the case of
the Feynman gauge, we can get the analytical form of
the equations
for the functions $\Gamma_1$ and $\Gamma_2$ directly from 
Eqs.~(\ref{eqg1feyn}) and~(\ref{eqg2feyn}). In the LC gauge the system of 
corresponding equations reads
\begin{eqnarray}
\label{eqg1lf}
\bar{u}(p_1)\Gamma_1u(p)
& = & {\displaystyle \bar{u}(p_1) g\gamma^{\nu}
(-\bar\Sigma_{\nu\rho}^{LC})\Gamma_2^{\rho}u(p)}
\nonumber\\
& + & {\displaystyle \bar{u}(p_1)\delta m'
\frac{(\sla{p}+\sla{\omega}\tau_1+m)}{m^2-M^2}\Gamma_1u(p)}
\nonumber\\
                           & + & {\displaystyle
\bar{u}(p_1)\left(Z\frac{m\sla{\omega}}
       {\omega\cd p}\right)
\frac{(\sla{p}+\sla{\omega}\tau_1+m)}{m^2-M^2}\Gamma_1u(p)}
       \nonumber\\
                           & + & {\displaystyle \bar{u}(p_1)g\delta
m'\left(-\frac{\sla{\omega}}
       {2 \omega\cd p}\right)\gamma^{\nu}
(-\bar\Sigma_{\nu\rho}^{LC})\Gamma_2^{\rho}u(p)} \nonumber\\
& + & {\displaystyle \bar{u}(p_1)\delta {m'}^2\left(-\frac{\sla{\omega}}
       {2 \omega\cd p}\right)
\frac{(\sla{p}+\sla{\omega}\tau_1+m)}{m^2-M^2}\Gamma_1u(p)},
\end{eqnarray}
\begin{eqnarray}
\label{eqg2lf}
\bar{u}(k_1)[\Gamma_2^{\mu}e_{\mu}^{\lambda}(k_2)]u(p)
                           & = & {\displaystyle
\bar{u}(k_1)g\gamma^{\mu}e_{\mu}^{\lambda}(k_2)
\frac{(\sla{p}+\sla{\omega}\tau_1+m)}{m^2-M^2}\Gamma_1u(p)}
\nonumber\\
  & + & {\displaystyle \bar{u}(k_1)
g^2\gamma^{\mu}e_{\mu}^{\lambda}(k_2)
                                  \left(-\frac{\sla{\omega}}{2 \omega\cd
p}\right)\gamma^{\nu}
(-\bar\Sigma_{\nu\rho}^{LC})\Gamma_2^{\rho}u(p)}\nonumber\\
                           & + & {\displaystyle \bar{u}(k_1)g\delta
m'\gamma^{\mu}e_{\mu}^{\lambda}(k_2)
       \left(-\frac{\sla{\omega}}
       {2 \omega\cd p}\right)
\frac{(\sla{p}+\sla{\omega}\tau_1+m)}{m^2-M^2}\Gamma_1u(p)}.
\end{eqnarray}
Note however that the polarization vectors $e^{\lambda}_{\mu}(k_2)$ 
here differ from those in the system~(\ref{eqg1feyn}),~(\ref{eqg2feyn}).
%


\subsection{Solution}\label{solut3}

As above, the structure of the one-body vertex function $\Gamma_1$ is
given by Eq.~(\ref{gamma1feyn}).
The number of independent invariant amplitudes for the vertex function
$\bar{u}(k_1)[\Gamma_2^{\mu}e_{\mu}^{\lambda}(k_2)]u(p)$ can be
calculated
analogously to the case
of the Feynman gauge, taking into account, however, that in the
LC gauge
the vector boson polarization vector is orthogonal to both $\omega$ and
the boson momentum. The
conditions
$e^{\lambda}(k_2) \cd \omega=0$ and
$e^{\lambda}(k_2)\cd k_2=0$ leave two independent components.
The total
number
of invariant amplitudes is thus $(2\times 2\times 2)/2=4$. We choose the
following set of invariant amplitudes:
\begin{equation}
\label{gamma2lf6}
\bar{u}(k_1)[\Gamma_2^{\mu}e_{\mu}^{\lambda}(k_2)]u(p)=
\bar{u}(k_1)\left(
b_1\gamma^{\mu}+
b_2\frac{m\sla{\omega}\gamma^{\mu}}{\omega\cd p}+
b_3\frac{p^{\mu}}{m}+
b_4\frac{m\sla{\omega}p^{\mu}}{\omega\cd p}\right)
e_{\mu}^{\lambda}(k_2)u(p).
\end{equation}
Again, in our approximation, the structures proportional to $p^{\mu}$
do not arise at all. We can therefore put $b_3=b_4=0$ and
reduce Eq.~(\ref{gamma2lf6}) to
\begin{equation}
\label{gamma2lf2}
\bar{u}(k_1)[\Gamma_2^{\mu}e_{\mu}^{\lambda}(k_2)]u(p)=
\bar{u}(k_1)\left(b_1\gamma^{\mu}+
b_2\frac{m\sla{\omega}\gamma^{\mu}}{\omega\cd p}
\right)e_{\mu}^{\lambda}(k_2)u(p).
\end{equation}
We insert Eqs.~(\ref{gamma1feyn}),~(\ref{gamma2lf2}) into
Eqs.~(\ref{eqg1lf}),~(\ref{eqg2lf})
and replace, as before, the spinors
$\bar{u}(p_1)$, $\bar{u}(k_1)$, $u(p)$
by $(\sla{p}_1+m)$, $(\sla{k}_1+m)$, $(\sla{p}+m)$, respectively,
and the polarization vectors $e_{\mu}^{\lambda}(k_2)$ by
$-[g_{\mu\mu'}-(k_{2\mu}\omega_{\mu'}+k_{2\mu'}\omega_{\mu})/(\omega\cd
k_2)]$.
After that we take the limit $M\to m$ and calculate the trace from
Eq.~(\ref{eqg1lf}).
Then we multiply
Eq.~(\ref{eqg2lf}) by $\gamma^{\mu'}$ and by
$m\sla{\omega}\gamma^{\mu'}/(\omega\cd p)$ successively and take traces again.
In calculating the traces we need to know the scalar products $p\cdot k_2$,
$k_1\cdot k_2$, and $\omega\cdot k_2$ in addition to those given by Eq.~(\ref{scprod1}).
These scalar products can be easily expressed through the variables
$s$ and $x$ [see Eq.~(\ref{scprod2})] by using the conservation law
$k_1+k_2=p+\omega\tau$ and the conditions $k_1^2=p^2=m^2$, $k_2^2=0$, $\omega^2=0$. 
The result is
\begin{equation}
\label{scprod3}
p\cd k_2=\frac{x(s-m^2)}{2},\quad k_1\cd k_2=\frac{s-m^2}{2},
\quad \frac{\omega\cd k_2}{\omega\cd p}=1-x.
\end{equation}

Finally, we get a system of three
homogeneous linear equations for the coefficients $a_1$, $b_{1,2}$.
It is given in Appendix~\ref{eqslc}.

Equating the determinant of this system to zero,
we come to a quadratic equation for $\delta m'$:
\begin{multline}\label{eq1lfb}
Det=\frac{2m^2(m^2-sx)(1+x^2)}{(1-x)}\left\{
\delta {m'}^2+2(2g^2A_1-m)\delta m' \right.\\
\left. +4g^2[g^2A_1^2-B_1(3g^2B_1+2g^2B_2-10g^2C_1-8g^2C_1'-Z)]
   \right.\\
\left.-2m(2g^2A_1-4g^2B_1-2g^2B_2+10g^2C_1+8g^2C_1'+Z)\vphantom{\delta {m'}^2}\right\}=0
\end{multline}
with the solution (we choose the minus sign at the square root, as explained
above)
\begin{eqnarray}
\label{delmlf} \delta m'&=&-2g^2A_1+m
-\sqrt{(m-2g^2B_1)[m-2(3g^2B_1+2g^2B_2-10g^2C_1-8g^2C_1'-Z)]}
\nonumber\\
&=& g^2A +m-  \sqrt{(m-g^2B)\left[m-g^2B- 2(g^2C - Z)\right]},
\end{eqnarray}
where we used the relations~(\ref{ctilde}) between the coefficients.
The substitution of Eq.~(\ref{delmlf}) into Eqs.~(\ref{eqF1lf})--(\ref{eq2lf})
allows to express the functions $b_{1,2}$ through $a_1$. In our two-body
approximation the functions $b_{1,2}$ are constants:
\begin{equation}
\label{b1lf}
b_1=2gma_1,
\end{equation}
\begin{eqnarray}
\label{b2lf} b_2&=&gma_1\left( 1-\sqrt{1-\frac{2(2g^2B_1+2g^2B_2-10g^2C_1-
8g^2C_1'-Z)}{m-2g^2B_1}} \right) \nonumber\\
&=& gm a_1\left(1- \sqrt{1-\frac{2(g^2C - Z)}{m-g^2B}}\right).
\end{eqnarray}
If we impose the condition
$Z-g^2C=0$
on the renormalization constant $Z$ which is still free, the function
$b_2$ turns to zero, as it should. Taking into account Eqs.~(\ref{dmprime}) 
and~(\ref{Zpp}), we find the following final solution:
\begin{equation}
\label{solgamma2lc}
b_1=2gma_1,\quad b_{2-4}=0,
\end{equation}
\begin{equation}
\label{soldelmlc} 
\delta m=g^2(A+B+C+C_{fc}+C_{bc}), \quad
Z_{\omega}=g^2(C+C_{fc}+C_{bc}).
\end{equation}
It formally differs from the corresponding 
solution~(\ref{solgamma2feyn})--(\ref{soldelmfeyn}) for the case of
the Feynman gauge by the presence of the boson contact
term $C_{bc}$. Note however that the values of $A$, $B$, $C$,
and $C_{fc}$ are different for the two gauges. Explicit
expressions for $\delta m$ and $Z_{\omega}$ are given in
Appendix~\ref{gboslc}. 

Though the solution for the state vector in the LC gauge,
Eqs.~(\ref{solgamma2lc}),
formally coincides with that obtained for the Feynman gauge 
and given by Eqs.~(\ref{solgamma2feyn}), the normalization factors
of the corresponding state vectors are different. These normalization 
factors are calculated fully analogously to how it was done
for the scalar boson case.

\section{Analytical results and discussion}\label{discuss}

Let us denote the contact term contribution by $C_c$:
\begin{equation}\label{CC}
C_c=\left\{
\begin{array}{ll} 
C_{fc} & \mbox{for scalar boson and for gauge
boson in the Feynman gauge},\\
C_{fc}+C_{bc} & \mbox{for gauge
boson in the LC gauge}.
\end{array}
\right.
\end{equation}
Then the expressions  for $Z_{\omega}$ and $\delta m$ in terms of
$A$, $B$, $C$, $C_{c}$ obtain the same form 
for all the cases discussed above:
\begin{eqnarray}\label{dmc}
\delta m &=&g^2(A+B+C+C_{c}),\\
\label{zoc}
Z_{\omega} &=&g^2(C+C_{c}).
\end{eqnarray}
The constants $A$, $B$, $C$, and $C_{c}$ are however different 
for different bosons. For the
scalar boson case all the constants are given in Appendix~\ref{scbos}. 
For the cases of
gauge boson in the Feynman and LC gauges they are given in 
Appendices~\ref{gbosf} and~\ref{gboslc}, respectively.
Substituting their explicit values into 
Eqs.~(\ref{dmc}) and~(\ref{zoc}), we find
the analytical formulas for $Z_{\omega}$ and $\delta m$.

For scalar boson:
\begin{eqnarray}
\delta m^{sc}&=&-\frac{g^2m}{8\pi^2}\log\frac{L^2}{m^2},
\label{dmsc}\\
Z_{\omega}^{sc}&=&-\frac{g^2m}{32\pi^2}\log\frac{L^2}{m^2} \label{zosc}.
\end{eqnarray}

For gauge boson in the Feynman gauge:
\begin{eqnarray}
\delta m^{F}&=&\frac{g^2m}{8\pi^2}\log\frac{L^2}{m^2},
\label{dmfr}\\
Z_{\omega}^{F}&=&-\frac{g^2m}{16\pi^2}\log\frac{L^2}{m^2} \label{zof}.
\end{eqnarray}

For gauge boson in the LC gauge:
\begin{eqnarray}
\delta m^{LC}&=&\frac{g^2m}{8\pi^2}\log\frac{L^2}{m^2},
\label{dmlc}\\
Z_{\omega}^{LF}&=&\frac{g^2m}{32\pi^2}\log\frac{L^2}{m^2} \label{zolc}.
\end{eqnarray}

Let us now discuss what happens with fermion-boson scattering amplitudes
calculated in CLFD within the two-body approximation. Although
this question, strictly speaking, goes beyond the scope of the present
paper, we think that our results are rather instructive and may be
useful for future investigations. 
In Sec.~\ref{need} we considered the perturbative expansion (up to terms
of order $g^4$) of such an amplitude for the scalar boson case and
showed that, after introducing the counterterm $Z_{\omega}$, it
became $\omega$-independent. Below we perform a similar analysis
of the scattering amplitudes without making expansions in powers of $g$.

In the two-body approximation the fermion-boson scattering amplitude $M$
can be obtained by sandwiching the fermion 2PGF, ${\cal D}(p)$, with the 
corresponding bispinors. The 2PGF, in its turn, is easily derived from 
Eq.~(\ref{solD}) by putting there $Z_2=1$ and by adding the 
counterterm~(\ref{Zom}) to $\delta m$. As a result, we find 
\begin{equation}
\label{amplnpert}
M=ig^2\bar{u}(p_1'){\cal D}(p)u(p_1)=
-g^2\bar{u}(p_1')\frac{\Gamma'(\sla{p}+m)\Gamma}{p^2-m^2+
\left[\delta m-{\cal M}^{(2)}(p)+Z_{\omega}
\left(\frac{m\sla{\omega}}{\omega\cd p}-1\right)\right]
(\sla{p}+m)}u(p_1),
\end{equation}
where $\Gamma=\Gamma'=1$ for scalar bosons, $\Gamma=\sla{e}(k_1)$,
$\Gamma'=\sla{e}(k_1')$ for gauge bosons, $k_1$ and $k_1'$ are the boson
four-momenta before and after scattering.
The superscript of the mass operator indicates that we are working in the 
two-body approximation. Using
Eq.~(\ref{CC}), we can write ${\cal M}^{(2)}(p)$ in the form
common for all boson types [cf. with Eq.~(\ref{massopabc})]:
\begin{equation}
\label{mtwobody}
{\cal M}^{(2)}(p)=g^2\left\{{\cal A}(p^2)+{\cal B}(p^2)\frac{\sla{p}}{m}
+[{\cal C}(p^2)+C_c]\frac{m\sla{\omega}}{\omega\cd p}\right\}.
\end{equation}
The coefficients $A$, $B$, and $C$ in Eqs.~(\ref{dmc}) and~(\ref{zoc})
are, respectively, 
the values of the functions ${\cal A}(p^2)$, ${\cal B}(p^2)$,
and ${\cal C}(p^2)$  at $p^2=m^2$.
In spite of that Eqs.~(\ref{amplnpert}) and~(\ref{mtwobody}) have the same 
form for all boson types, these functions, as well as the constants
$\delta m$, $C_c$, and $Z_{\omega}$, are different for different bosons.

Substituting Eq.~(\ref{mtwobody}) into Eq.~(\ref{amplnpert}) 
and using Eqs.~(\ref{dmc}) and~(\ref{zoc}), we see
that the $\omega$-dependence of the amplitude is determined by the
term 
\begin{equation}
\label{omdep}
\left(g^2\left[{\cal C}(p^2)+C_c\right]
-Z_{\omega}\right)\frac{m\sla{\omega}}{\omega\cd p}
(\sla{p}+m)=g^2\left[{\cal C}(p^2)-{\cal C}(m^2)\right]
\frac{m\sla{\omega}}{\omega\cd p}(\sla{p}+m)
\end{equation}
in the denominator of Eq.~(\ref{amplnpert}). In Appendix~\ref{scbos}
we proved that ${\cal C}(p^2)-{\cal C}(m^2)=0$ for the scalar boson case.
For the case of gauge boson in the Feynman gauge the situation is
completely the same, since the function ${\cal C}(p^2)$ differs from
that for scalar bosons, given by Eq.~(\ref{abccalC}), 
by a factor of 2 only (see the argumentation in Appendix~\ref{gbosf}).
So, the quantity~(\ref{omdep}) equals zero for both 
scalar bosons and gauge bosons in the Feynman gauge, and the 
scattering amplitude~(\ref{amplnpert}) does not depend on $\omega$ at all.

The latter statement is not true for the case of the LC gauge. The function
${\cal C}(p^2)$ can be obtained by substituting the integral 
from Eq.~(\ref{senflc}), without putting $p^2=m^2$, into 
the last of Eqs.~(\ref{eq4sen}). Technical details are completely
equivalent to those exposed in Appendix~\ref{scbos}. Thus, we get
for the LC gauge:
\begin{equation}
\label{ccallc}
{\cal C}(p^2)=-\frac{1}{16\pi^2m}\int_0^{L^2}dk_{\perp}^2
\int_{\delta}^{1-\delta} dx\,\frac{k_{\perp}^2(x^2-2x+2)+m^2x^2-p^2x^2(1-x)^2}
{x^2(1-x)[k_{\perp}^2+m^2x-p^2x(1-x)]}.
\end{equation}
Subtracting from Eq.~(\ref{ccallc}) the value ${\cal C}(m^2)$, we
arrive at the expression
\begin{equation}
{\cal C}(p^2)-{\cal C}(m^2)=\frac{(p^2-m^2)}{16\pi^2m}\int_0^{L^2}dk_{\perp}^2\int_{\delta}^1
dx\,\frac{k_{\perp}^2(2x^2-3x+2)+m^2x^3}
{x[k_{\perp}^2+m^2x^2][k_{\perp}^2+m^2x-p^2x(1-x)]}.
\label{dclc}
\end{equation}
The latter integral diverges logarithmically at $x=0$ and can not be zero.
From here follows that the scattering amplitude~(\ref{amplnpert}) for 
gauge bosons
in the LC gauge, generally speaking, depends on $\omega$. 
Such an unpleasant feature is however not specific to LFD, 
but also takes place in 4DF formalism, 
as explained in Ref.~\cite{bass}. 
This is caused by the bad regularization of the singularity 
in $\omega \cd k$ to which the function ${\cal C}(p^2)$ is sensitive.

Now we can summarize the results obtained above.
\begin{itemize}
\item LFD requires an additional counterterm in the
Hamiltonian,
depending on the light front orientation (i.~e., on the four-vector $\omega$).
In the two-body approximation this counterterm in momentum space has
the form
$-Z_{\omega}\frac{m\sla{\omega}}{\omega\cd p}$. 
In the nonperturbative calculations presented above it eliminates the
$\omega$-dependence of the two-body wave function residue at the pole
$s=m^2$. 

\item
We have found, for the truncation $|F\rangle + |Fb\rangle$, the
true nonperturbative solutions for the mass counterterm $\delta m$ 
and the composite fermion state vector for the cases of scalar and gauge bosons.  
Expanded in a series in powers of the coupling constant $g$, these solutions
contain all degrees of the latter. Besides that, the two-body wave functions 
have a non-zero spin components explicitly depending on $\omega$. 
The additional counterterm, $Z_{\omega}$, eliminates these components.
After that 
the solutions~(\ref{dmsc})--(\ref{zolc}) coincide with the
perturbative ones, as they should if 
the $|F\rangle + |Fb\rangle$ Tamm-Dancoff truncation is used.
A real difference between perturbative and nonperturbative results
must appear when higher Fock sectors are taken
into account. In the latter case one can hardly expect the vertex functions to be constants.
Consequently, there is no any reason to demand that all their $\omega$-dependent components
equal zero identically in order to get an $\omega$-independent residue 
of the two-body wave function at its pole.

\item
The mass shift $\delta m$ is gauge invariant, as is seen from the
coincidence of Eqs.~(\ref{dmfr}) and~(\ref{dmlc}).

\item
The mass shift $\delta m$ does not coincide with the corresponding value found 
in Ref.~\cite{must91}, since it depends on the choice of cutoffs. 
Our cutoffs are different from those used in Ref.~\cite{must91}. 
We show in Appendix~\ref{cutoffs} that our formulas with the
cutoffs taken from Ref.~\cite{must91} reproduce $\delta m$
obtained there.

\item
The value of the counterterm $Z_{\omega}$ depends on gauge, as expected.

\item
In spite of the fact that intermediate calculations involve strongly divergent
integrals [see, e.~g., Eqs.~(\ref{Ac}), (\ref{Afc}), 
(\ref{Cf}), (\ref{Cfc}), (\ref{eq5bC})--(\ref{eq5bCbc})], the final
results for $\delta m$ and $Z_{\omega}$ contain logarithmic divergences only.

\end{itemize}


\section{Conclusion}\label{concl}

We considered a spin 1/2 fermion state consisting of the bare fermion 
$F$ coupled to bosons $b$
with spin either zero or one, in the approximation where the first two
Fock sectors ($|F\rangle+|Fb\rangle$) are kept in the state vector.
The spin one bosons are the gauge ones.

Although the state vector of the composite fermion is simplified, the study
of this model is of utmost importance in order to settle the various
pieces of the light-front Hamiltonian we start from. While the light-front Hamiltonian is
simple at the tree level, this is not true anymore when higher order processes
are considered. Due to the explicit covariance of our approach, we exhibit
in the present study the exact operator structure of the counterterms needed for the renormalization. 
In the two-body approximation two counterterms are required to renormalize
the theory: a mass counterterm and a specific one for LFD, which explicitly depends on the
orientation of the light front plane (i.~e. on the four-vector $\omega$). 
We calculated them in an explicit form. While the mass counterterm is analogous to that which
appears in standard 4DF approach, the specific LFD counterterm has no such an analogue.

These two counterterms generate additional contributions to the light-front Hamiltonian,
due to their coupling with
contact interactions. The latters also depend on the light front plane
orientation. The solutions we found for
the couplings of scalar and gauge bosons to a spin $1/2$ fermion are in
complete agreement with the CLFD perturbation theory, although our
calculation does not rely on perturbative expansions. The condition
imposed on the mentioned above specific counterterm eliminates the 
$\omega$-dependence of the two-body wave
function residue at its pole. 

We explicitly showed that the mass counterterm $\delta m$ is gauge independent.
We calculated the
state vector in two gauges --- the Feynman and the LC ones. Though
finally the number of the two-body wave function spin components turned out to be the same in both
gauges, this is a peculiarity of the $|F\rangle+|Fb\rangle$ Fock space
truncation. Such a property should disappear when 
higher Fock components are taken into account. 
At the same time, physical observables (e.~g, the electromagnetic form 
factors) must be the same in any gauge.

The results reported here are encouraging in the perspective of tackling physical
composite systems in QED and QCD. This will be a subject of
forthcoming researches.


\section*{Acknowledgements}

Two of us (V.A.K. and A.V.S.) are sincerely grateful for the warm
hospitality
of Laboratoire de Physique Corpusculaire, Universit\'e Blaise Pascal, in
Clermont-Ferrand, where the present study was performed. This work is partially
supported by the French-Russian PICS and  RFBR grants Nos. 1172 and
01-02-22002 as well as by the RFBR grant No. 02-02-16809.


\appendix

\section{Light-front Hamiltonians}\label{lfh}

\subsection{Scalar boson}\label{scalbos}

For non-zero spin fields, the light-front Hamiltonian should
incorporate the so-called contact interactions. Their origin and
derivation are explained, e.g., in Ref.~\cite{bpp} (in the LC
gauge, for the gluon field). For completeness, we  briefly explain
this derivation, in the Feynman gauge too.

We start with the case of scalar boson field. The corresponding
field theory Lagrangian, at the tree level, is
\begin{equation}
\label{lagr} 
{\cal L}=\frac{i}{2}\left[\bar{\Psi}\gamma^{\nu}(\partial_{\nu} \Psi)-
(\partial_{\nu}\bar{\Psi})\gamma^{\nu} \Psi\right]-
m\bar{\Psi}\Psi+\frac{1}{2}\left[\partial_{\nu}\Phi
\partial^{\nu}\Phi-\mu^2\Phi^2\right]+
g\bar{\Psi}\Psi\Phi,
\end{equation}
where $\Psi$ and $\Phi$ are, respectively, the Heisenberg spinor
and scalar field operators.
In a standard way we obtain the energy-momentum tensor
\begin{equation}
\label{enmomtensor}
\Theta^{\nu\rho}=
\partial^{\rho}\bar{\Psi}\frac{\partial {\cal L}}{\partial
(\partial_{\nu}\bar{\Psi})}+
\frac{\partial {\cal L}}{\partial
(\partial_{\nu}\Psi)}\partial^{\rho}\Psi+
\frac{\partial {\cal L}}{\partial
(\partial_{\nu}\Phi)}\partial^{\rho}\Phi-
g^{\nu\rho}{\cal L}
\end{equation}
and the four-momentum operator
\begin{equation}
\label{momoperator}
{\cal P}^{\rho}=\int d\sigma_{\nu}(x)\Theta^{\nu\rho}.
\end{equation}
Integration in Eq.~(\ref{momoperator}) is performed on the
three-dimensional
space element orthogonal to the "time" direction.
The role of time is played in CLFD by the invariant
combination $\omega\cd x$. To simplify the subsequent algebraical
manipulations it is
convenient to introduce temporarily a particular
value\footnote{In principle, one can carry out the derivation
without
specifying the particular four-vector $\omega=(1,0,0,-1)$, but
introducing
four orthogonal four-vectors, one of which is  $\omega$.} of the
four-vector
$\omega$,
namely, $\omega=(1,0,0,-1)$, corresponding to the usual form of
noncovariant LFD on the plane $x^0+x^3=0$. In the final results
we will return to explicitly covariant notations. This is simply achieved
by the replacement of the plus-component of any four-vector $a$ by the
contraction $\omega\cd a$.

Let us introduce also the light-front coordinates
\begin{equation}
\label{lfcoord} x^+=x^0+x^3,\quad x^-=x^0-x^3,\quad {\bf
x}^{\perp}=\{x^1,x^2\}.
\end{equation}
The scalar product is defined as
\begin{equation}
\label{sp}
a_{\nu}b^{\nu}=\frac{1}{2}a^+b^-+\frac{1}{2}a^-b^+-
{\bf a}^{\perp}\cdot {\bf b}^{\perp}.
\end{equation}
The metric tensor is
\begin{equation}
\label{metrictensor}
g^{+-}=g^{-+}=2\,\,\,\,\,g_{+-}=g_{-+}=\frac{1}{2}\,\,\,\,\,
g^{11}=g^{22}=g_{11}=g_{22}=-1,
\end{equation}
all the other components being zeroes. In such a formalism
the ${\cal P}^+$ and ${\bg{\cal P}}^{\perp}$
components of the momentum operator are free, while the ${\cal P}^-$
component contains interaction.
${\cal P}^-$ is given by the expression
\begin{equation}
\label{momminus}
{\cal P}^-=\frac{1}{2}\int d^2x^{\perp}dx^-\Theta^{+-}.
\end{equation}
 From Eqs.~(\ref{enmomtensor}) and~(\ref{momoperator}) with the
Lagrangian~(\ref{lagr}) we get
\begin{equation}
\label{thetapm}
{\cal P}^{-}=\frac{1}{2}\int d^2x^{\perp}dx^-\left[-i\bar{\Psi}\gamma^-
\partial^+
\Psi+2i\bar{\Psi}\bg{\gamma}^{\perp}\cdot \bg{\partial}^{\perp}
\Psi+2m\bar{\Psi}\Psi +(\bg{\partial}^{\perp}\Phi)^2+\mu^2\Phi^2
-2g\bar{\Psi}\Psi\Phi\right].
\end{equation}

The time evolution of the operators $\Psi$ and $\Phi$ is governed by the
equations of motion obtained from the Lagrangian~(\ref{lagr}). Thus
we have the following Dirac-type equation for $\Psi$:
\begin{equation}
\label{eqmot1} (i \sla{\partial}-m)\Psi=-g\Phi\Psi.
\end{equation}
The spinor field $\Psi$ has four components. However, it
turns out that Eq.~(\ref{eqmot1}) in the light-front
coordinates results in a time-dependent
equation only for two components among the four. For the other two
components, it gives a
time-independent constraint. The elimination of these two components from
the Hamiltonian, by using this constraint, generates
an extra term in the Hamiltonian, which is just the contact interaction.

To demonstrate this in more detail, let us split
the four-component spinor $\Psi$ into two
two-component pieces, $\Psi^{(+)}$ and $\Psi^{(-)}$,
defined as
\begin{equation}
\label{psipm}
\Psi^{(\pm)}=\Lambda^{(\pm)}\Psi,
\end{equation}
where $\Lambda^{(\pm)}=\frac{1}{4}\gamma^{\mp} \gamma^{\pm}$
with $\gamma^{\pm}=\gamma^0\pm \gamma^3$
are projection operators with the properties
$$
{\Lambda^{(\pm)}}^2=\Lambda^{(\pm)},\,\,\,\,\Lambda^{(+)}\Lambda^
{(-)}=\Lambda^{(-)}\Lambda^{(+)}=0,
\,\,\,\,{\Lambda^{(\pm)}}^{\dag}=\Lambda^{(\pm)},
$$
$$
\Lambda^{(\pm)}\gamma^0=\gamma^0\Lambda^{(\mp)},\,\,\,\,
\Lambda^{(\pm)}\bg{\gamma}^{\perp}=\bg{\gamma}^{\perp}\Lambda^{(\pm)}.
$$
The equation of motion~(\ref{eqmot1}) also splits into two equations
\begin{equation}
\label{eqpsip} i \partial^- \Psi^{(+)}-(i{\bg{\alpha}}^{\perp}\cd
{\bg{\partial}}^{\perp}+ m\gamma^0 -g\Phi\gamma^0)\Psi^{(-)}=0,
\end{equation}
\begin{equation}
\label{eqpsim} i \partial^+ \Psi^{(-)}-(i{\bg{\alpha}}^{\perp}\cd
{\bg{\partial}}^{\perp}+ m\gamma^0 -g\Phi\gamma^0)\Psi^{(+)}=0.
\end{equation}
It is important that Eq.~(\ref{eqpsim}) does not contain the "time"
derivative $\partial^-\equiv 2\frac{\partial}{\partial x^+}$ and,
hence, represents a
constraint connecting the components $\Psi^{(+)}$ and $\Psi^{(-)}$ at
any "time":
\begin{equation}
\label{connect} \Psi^{(-)}=\frac{1}{i\partial^+}(i{\bg{\alpha}}^{\perp}\cd
{\bg{\partial}}^{\perp}+ m\gamma^0 -g\Phi\gamma^0)\Psi^{(+)}.
\end{equation}
The Hermitian integral operator $\frac{1}{i\partial^+}$
acts on a  function $f$ according to
the following rule:
\begin{equation}
\label{operator}
\frac{1}{i\partial^+}f(x^-)=-\frac{i}{4}\int dy^-\epsilon(x^--y^-)f(y^-)\
,
\end{equation}
where $\epsilon(x)=1$ for $x>0$ and $\epsilon(x)=-1$ for $x<0$.
In momentum space it corresponds to the division by $p^+$
(or by $\omega\cd p$), where $p$ is the four-momentum conjugated to
the coordinate $x$.

We represent the state vector in Fock space formed by the action
of the usual (free) creation operators on the vacuum. One should also
express the Hamiltonian in
terms of these free operators, i.e., the operators in Schr\"odinger
representation. The momentum operator $P^-$ in Schr\"odinger
representation has the same form as ${\cal P}^{-}$, changing Heisenberg
field operators
to Schr\"odinger ones. The Heisenberg operators $\Psi^{(+)}$ and
$\Phi$ are connected with the corresponding Schr\"{o}dinger operators
$\psi^{(+)}$ and $\varphi$ by the relations
\begin{equation}
\label{represent}
\Psi^{(+)}=e^{\frac{1}{2}{\cal P}^{-}x^+}\psi^{(+)}
e^{-\frac{1}{2}{\cal P}^{-}x^+},\,\,\,\,
\Phi=e^{\frac{1}{2}{\cal P}^{-}x^+}
\varphi e^{-\frac{1}{2}{\cal P}^{-}x^+},
\end{equation}
where it is supposed that at $x^+=0$ the Heisenberg and Schr\"{o}dinger
operators coincide. As it can be seen from Eq.~(\ref{represent}),
the possibility to express the momentum operator
through the free fields is essentially based on the existence of a "time"
moment
(chosen to be $x^+=0$), when all field components entering this operator
are free simultaneously.
However, in CLFD the constraint~(\ref{connect}) connecting
the spinor field components $\Psi^{(+)}$ and $\Psi^{(-)}$ at {\em any}
time moment involves interaction terms.
By this reason in order to construct the momentum operator one must
express the latter through independent field components only and after
that substitute the Heisenberg field operators by the Schr\"odinger ones.
In the case considered above the independent components
are $\Psi^{(+)}$ and $\Phi$.

So, changing in Eq.~(\ref{thetapm}) $\Psi^{(+)}$ by  $\psi^{(+)}$ and
$\Phi$ by  $\varphi$, denoting
$$
\psi=\left[1+\frac{i{\bg{\alpha}}^{\perp}\cd {\bg{\partial}}^{\perp}
+\gamma^0 m}{i\partial^+}\right]\psi^{(+)},
$$
where $\psi$ satisfies the free Dirac equation, and returning
to the covariant
notations, we find for the interaction part of the four-momentum:
\begin{equation}
\label{pintsc}
P^{int}_{\rho}=\omega_{\rho}\int\delta(\omega\cd x)d^4x
\left[-g\bar{\psi}\psi\varphi
+g^2\bar{\psi}\varphi\frac{\sla{\omega}}{2i(\omega\cd\partial)}
\varphi\psi\right].
\end{equation}
The integrand in Eq.~(\ref{pintsc}) is nothing else than the interaction
CLFD Hamiltonian.

To get the complete Hamiltonian incorporating the mass renormalization and 
the specific CLFD counterterm discussed in Sec.~\ref{need} [see
Eq.~(\ref{Zom})] we should
add to Eq.~(\ref{lagr}) the following additional terms:
\begin{equation}
\label{lagrc}
\delta m\bar{\Psi}\Psi+Z_{\omega}\bar{\Psi}
\left(\frac{m\sla{\omega}}{i\omega\cd \partial}-1\right)\Psi
\equiv \delta m'\bar{\Psi}\Psi+
Z_{\omega}\bar{\Psi}\frac{m\sla{\omega}}{i\omega\cd\partial}\Psi,
\end{equation}
where $\delta m'$ is defined by Eq.~(\ref{dmprime}).
It is easy to see that the counterterm with $Z_{\omega}$ in Eq.~(\ref{lagrc})
just generates Eq.~(\ref{Zom}) in momentum representation.

A question may arise whether it makes sense to add the counterterm
with $Z_{\omega}$, explicitly depending on the light-front position,
to the Lagrangian. On the one hand, the Lagrangian as a fundamental
quantity describing the physical system considered may "know" nothing
about CLFD and truncated Fock space. 
On the other hand, we can not neglect the fact that the structure of 
counterterms
substantially depends on the method of calculations. Since CLFD is
based on Hamiltonian dynamics, and the Hamiltonian is derived from
the Lagrangian, we can incorporate necessary counterterms, formally
including the latters directly in the initial Lagrangian. At this level such
a procedure looks like a formal trick allowing to simplify the
construction of the Hamiltonian.

After adding the counterterms~(\ref{lagrc}) we need not repeat the derivation
of the Hamiltonian from the very beginning. In a first step, we put 
$Z_{\omega}=0$ on the r.~h.~s. of Eq.~(\ref{lagrc}) and take into 
account the term with $\delta m'$ only. The
additional
term $\delta m' \bar{\Psi}\Psi$ in the Lagrangian changes the equation of
motion~(\ref{eqmot1}) to
\begin{equation}
\label{eqmotnew}
(i \sla{\partial} - m)\Psi=-(g\Phi+\delta m')\Psi.
\end{equation}
Note that Eq.~(\ref{eqmotnew}) can be obtained from Eq.~(\ref{eqmot1})
by the
simple substitution $\Phi \to \Phi+\delta m'/g$. Hence, making an
analogous
substitution $\varphi \to \varphi+\delta m'/g$ in Eq.~(\ref{pintsc}),
we find the mass operator including mass renormalization effects.
In a second step, we add  the counterterm
$Z_{\omega}\bar{\Psi}[m\sla{\omega}/i(\omega\cd\partial)]\Psi$ to the Lagrangian.
Since $\bar{\Psi}\sla{\omega}\Psi=2{\Psi^{(+)}}^{\dag}\Psi^{(+)}$,
this counterterm involves the $\Psi^{(+)}$-components only. By
this reason, it does not affect the constraint~(\ref{connect}) and
comes into the Hamiltonian as an addendum, without changing its 
initial form. 

Summarizing, we conclude that the final momentum operator including
the whole set of counterterms can be obtained from Eq.~(\ref{pintsc})
by changing $\varphi\to \varphi+\delta m'/g$ and by
adding to the resulting integrand the quantity 
$-Z_{\omega}\bar{\psi}[m\sla{\omega}/i(\omega\cd\partial)]\psi$.
After simple algebraic transformations we get the regularized momentum
operator
\begin{equation}
\label{hamiltonianap}
P^{int}_{\rho}=\omega_{\rho}\int d^4x\delta(\omega x)\, H^{int}(x),
\end{equation}
with  $H^{int}(x)$ given by Eq.~(\ref{hamiltonian}).
Except for the "normal" interaction $-g\bar{\psi}\psi\varphi$,
all other terms
in the Hamiltonian~(\ref{hamiltonian}) are contact ones.


\subsection{Vector boson}\label{fg}

The Lagrangian of a system of interacting spinor $\Psi$
and massless vector $\Phi^{\mu}$ fields is, also at the tree level,
\begin{equation}
\label{lagrvph}
{\cal L}=\frac{i}{2}\left[\bar{\Psi}\gamma^{\nu}(\partial_{\nu} \Psi)-
(\partial_{\nu}\bar{\Psi})\gamma^{\nu} \Psi\right]-
m\bar{\Psi}\Psi
-\frac{1}{4}F^{\mu\nu}F_{\mu\nu}-
\frac{\lambda}{2}(\partial\cdot\Phi)^2+
g\bar{\Psi}\sla{\Phi}\Psi,
\end{equation}
where $F^{\mu\nu}=\partial^{\mu}\Phi^{\nu}-\partial^{\nu}\Phi^{\mu}$.
We have added to the Lagrangian a term proportional
to the square of the vector field divergence in order to have a
possibility to choose different gauges.
Due to the gauge invariance all physical results must be independent of
the choice of $\lambda$.

The equations of motion read:
\begin{equation}
\label{eqmot1v}
(i \sla{\partial} - m)\Psi=-g\sla{\Phi}\Psi,
\end{equation}
\begin{equation}
\label{eqmot2v}
\partial^2\Phi^{\nu}-(1-\lambda)\partial^{\nu}(\partial\cdot\Phi)=
-g\bar{\Psi}\gamma^{\nu}\Psi,
\end{equation}
where $\partial^2\equiv \partial\cdot\partial=\partial^{+}\partial^{-}-
{\bg{\partial}^{\perp}}^2$.
The momentum operator acquires the form
\begin{eqnarray}
{\cal P}^{-}&=&\frac{1}{2}\int d^2x^{\perp}dx^-
\left\{-i\bar{\Psi}\gamma^-
\partial^+
\Psi+2i\bar{\Psi}\bg{\gamma}^{\perp}\cdot
\bg{\partial}^{\perp} \Psi+2m\bar{\Psi}\Psi\right. \nonumber \\
&&+\frac
{1-\lambda}{4}\left[(\partial^-\Phi^+)^2-(\partial^+\Phi^-)^2\right]+
\lambda (\bg{\partial}^{\perp}\cdot\bg{\Phi}^{\perp})^2
+[\bg{\partial}^{\perp}\times \bg{\Phi}^{\perp}]^2 \nonumber \\
\label{thetapmvph}
&&\left.
-(\bg{\partial}^{\perp}\Phi^-)\cdot(\bg{\partial}^{\perp}\Phi^+)
+(1-\lambda)(\bg{\partial}^{\perp}\Phi^-)\cdot
(\partial^+\bg{\Phi}^{\perp})
-2g\bar{\Psi}\sla{\Phi}\Psi\right\}.
\end{eqnarray}
Analogously to the scalar boson case, only two spinor components coming
into $\Psi^{(+)}$ are independent, $\Psi^{(-)}$ being expressed through
$\Psi^{(+)}$. To demonstrate it we split, as before,
the spinor $\Psi$ into $\Psi^{(+)}$ and
$\Psi^{(-)}$. Instead of Eq.~(\ref{eqpsim}) we now have
\begin{equation}
\label{eqpsimv}
(i \partial^++g\Phi^+)\Psi^{(-)}-(i\bg{\alpha}^{\perp}\cdot
\bg{\partial}^{\perp}+ m\gamma^0 - g
\gamma^0\Lambda^{(+)}\sla{\Phi})\Psi^{(+)}=0.
\end{equation}
Note that since $\Lambda^{(+)}\sla{\Phi}\Psi^{(+)}=
-\bg{\gamma}^{\perp}\cdot\bg{\Phi}^{\perp}\Psi^{(+)}$,
the $\Phi^{-}$ component does not contribute to
Eq.~(\ref{eqpsimv}) at all.
The solution of this equation is
\begin{equation}
\label{connectv}
\Psi^{(-)}\equiv \Psi_0^{(-)}+\Upsilon_1^{(-)}+\Upsilon_2^{(-)},
\end{equation}
\begin{equation}
\label{kinv}
\Psi_0^{(-)}=\frac{1}{i\partial^+}(i\bg{\alpha}^{\perp}\cdot
\bg{\partial}^{\perp}+m\gamma^0)\Psi^{(+)},
\end{equation}
\begin{equation}
\label{dinv}
\Upsilon_1^{(-)}=-
\frac{g}{i\partial^+}\gamma^0\Lambda^{(+)}\sla{\Phi}\Psi^{(+)},\,\,\,\,
\Upsilon_2^{(-)}=\frac{1}{i{\cal D}^+}
(i\bg{\alpha}^{\perp}\cdot\bg{\partial}^{\perp}+m\gamma^0
-g\gamma^0\Lambda^{(+)}\sla{\Phi})\Psi^{(+)},
\end{equation}
where the Hermitian operator $\frac{1}{i{\cal D}^+}$ is defined as
\begin{equation}
\label{operatorD}
\frac{1}{i{\cal D}^+}f(x^-)=-\frac{i}{4}\int dy^-\epsilon(x^--y^-)\left[
\exp\left(\frac{ig}{2}\int_{y^-}^{x^-}\Phi^+(z^-)dz^-\right)-1\right]f(y^-
).
\end{equation}
The operator $\frac{1}{i{\cal D}^+}$ is constructed so that it turns to
zero for any $f(x^-)$ if $g=0$ or $\Phi^+=0$.

As far as the vector field $\Phi^{\nu}$ is concerned, the situation
differs strongly from that for scalar bosons. Indeed, the vector field
has not one, but four components satisfying Eq.~(\ref{eqmot2v}).
In principle, some
restrictions can be imposed on these components, reducing the number
of independent ones (this just corresponds to a certain choice of gauge).
Besides that, the momentum
operator~(\ref{thetapmvph}) depends
on the first order "time" derivative $\partial^-\Phi^+$,
while Eq.~(\ref{thetapm}) does
not depend on $\partial^-\Phi$. When constructing
the momentum operator in Schr\"odinger representation, we can not
merely substitute $\partial^-\Phi^+$ by the corresponding free
field derivative $\partial^-\varphi^+$. The reason for this is that
$\partial^-\Phi^+$ is not an independent quantity, because
Eq.~(\ref{eqmot2v}) taken for $\nu=+$ allows to express
$\partial^-\Phi^+$ through the other field components and their "spatial"
derivatives:
\begin{equation}
\label{dmin}
\partial^-\Phi^+=\left( \frac{2}{1+\lambda} \right)
\left[ \frac{1}{\partial^+} \left(
{{\mbox{\boldmath
$\partial$}}^{\perp}}^2\Phi^+-2g{\Psi^{(+)}}^{\dag}\Psi^{(+)}
                              \right)
+\left( \frac{1-\lambda}{2} \right)
(\partial^+\Phi^--2\bg{\partial}^{\perp}\cdot\bg{\Phi}^{\perp})
\right].
\end{equation}
It is important that this connection includes interaction terms.

Applying the operator $\partial_{\nu}$ to the both sides
of Eq.~(\ref{eqmot2v}) gives
\begin{equation}
\label{cond1}
\lambda\partial^2(\partial\cdot \Phi)=0.
\end{equation}
In deriving Eq.~(\ref{cond1}) we took into account the current
conservation, $\partial_{\nu}(\bar{\Psi}\gamma^{\nu}\Psi)=0$,
which, in its turn, follows from Eq.~(\ref{eqmot1v}). In spite of that,
Eq.~(\ref{cond1})
contains vector field components only and cannot be regarded as a
constraint because it involves "time" derivatives of the second order.


\subsubsection{Feynman gauge}\label{fgauge}

In the Feynman gauge all the four vector field components are treated as
being independent. The light-front free vector boson propagator is
$-g_{\mu\nu}\delta(q^2)\theta(\omega q)$. To meet these requirements
one should put $\lambda=1$. Indeed, consider the solution
of Eq.~(\ref{eqmot2v}) corresponding to a free boson (i.~e. at $g=0$).
This solution is
a superposition of plane waves $\varphi^{\nu}(q)e^{-iqx}$ with $q^2=0$.
Substituting it into Eq.~(\ref{eqmot2v}), one gets $(1-\lambda)q^{\nu}
[q\cdot\varphi(q)]=0$. If $q\cdot\varphi(q)=0$, only three
field components are independent, but not four. So, we are forced to put
$\lambda=1$. Under this condition, the momentum
operator~(\ref{thetapmvph}) does not depend on $\partial^-\Phi^+$.

Since in the Feynman gauge the independent fields are $\Psi^{(+)}$
and $\Phi^{\nu}$, the subsequent analysis is quite similar to that
made for the scalar boson case. It leads to the following expression for
the interaction part of the momentum operator in Schr\"odinger
representation:
\begin{equation}
\label{pminintvphf}
P^{int}_{\rho}=\omega_{\rho}\int d^4x\delta(\omega x)\, H^{int}_F(x)
\end{equation}
with the interaction Hamiltonian
\begin{equation}
\label{feynman}
H^{int}_F(x)=-g\bar{\psi}\sla{\varphi}\psi
+g^2\bar{\psi}\sla{\varphi}\frac{\sla{\omega}}
{2i(\omega\cdot\partial)}\sla{\varphi}\psi
+g^2\bar{\psi}\sla{\varphi}
\frac{\sla{\omega}}{2i(\omega\cdot D)}\sla{\varphi}\psi.
\end{equation}
The operator $1/(\omega\cdot D)$ is given by the r.~h.~s.
of Eq.~(\ref{operatorD}) after the
replacement $\Phi^+\to \varphi^+\equiv \omega\cdot\varphi$. In an
explicitly covariant form it can be rewritten as
\begin{equation}
\label{operatorDexp}
\frac{1}{i(\omega\cd D)}f=\left[R^*\frac{1}{i(\omega\cd\partial)} R
-\frac{1}{i(\omega\cd\partial)}\right]f,
\end{equation}
where
$R=\exp\left[\frac{g}{i\omega\cdot\partial}(\omega\cdot\varphi)\right]$.
Note that the operator $\frac{g}{i(\omega\cd\partial)}$ inside the
exponent
acts on $\omega\cd\varphi$ only, while the other such operators
in Eq.~(\ref{operatorDexp}) act on all functions standing to the right of
them.

The second term in Eq.~(\ref{feynman}) is a fermion contact term.
In the Feynman
gauge, to  order $g^2$, contact terms related to the
massless vector
field are absent. The decomposition of the last term on the r.~h.~s. of
Eq.~(\ref{feynman})
starts with $g^3$ and contains an infinite number of terms. The matrix
element of
a term of arbitrary order contains the contractions of the vector field
operators entering Eq.~(\ref{feynman}) between themselves and with the
external
fields. As is shown in Sec.~\ref{ctfeyn}, 
in the Feynman gauge the internal contractions
disappear since they are always proportional to $\omega^2=0$. The
contractions with
external fields only survive. Therefore, in truncated Fock space, the
number of contact terms is finite and  the Hamiltonian~(\ref{feynman})
has polynomial dependence on the coupling constant and field operators.
We emphasize
that this property is a merit of the Feynman gauge, while another choice
of gauge may lead to appearance of an infinite number of contact terms
even
for truncated Fock space.

Incorporating the self-interaction $\delta m'\bar{\Psi}\Psi$ by means of the
substitution
$\sla{\varphi}\to\sla{\varphi}+\delta m'/g$
[with $\delta m'$ given by Eq.~(\ref{dmprime})]
and adding the counterterm 
$-Z_\omega\bar{\psi}[m\sla{\omega}/i(\omega\cd \partial)]\psi$, we obtain
the Hamiltonian~(\ref{hamiltonian_f}) with $H'(x)$ given by
Eqs.~(\ref{hamfeyn})--(\ref{hamfeyn3}).


\subsubsection{Light-cone gauge}\label{lcg}

Maximal simplifications are obtained in the LC gauge where
$\Phi^+=0$ (equivalent to $\omega\cdot\Phi=0$ in CLFD). In this
gauge the bosonic propagator
$-[g_{\mu\nu}-(q_{\mu}\omega_{\nu}+q_{\nu}\omega_{\mu})/(\omega\cdot
q)] \delta(q^2)\theta(\omega\cdot q)$ is transversal to $\omega$
in both indices. Note that expressions for vector meson
propagators in the LC gauge can be found in Refs.~\cite{bpp,cdkm}.

For $\Phi^+=0$, we get from Eq.~(\ref{dmin}):
\begin{equation}
\label{kindinlf}
\Phi^-=\frac{2}{{\partial^+}}
(\bg{\partial}^{\perp}\cd \bg{\Phi}^{\perp})+
\frac{4g}
{1-\lambda}\frac{1}{{\partial^+}^2}{\Psi^{(+)}}^{\dag}\Psi^{(+)}.
\end{equation}
This is a constraint indicating that only two transversal components
$\bg{\Phi}^{\perp}$ among four initial ones, $\Phi^{\nu}$,
are independent.
Under these requirements the vector field, generally speaking,
does not satisfy the Lorentz condition:
\begin{equation}
\label{diverglf}
\partial\cd\Phi=\frac{1}{2}\partial^{+}\Phi^{-}-
\bg{\partial}^{\perp}\cd \bg{\Phi}^{\perp}=
\frac{2g}
{1-\lambda}\frac{1}{\partial^+}{\Psi^{(+)}}^{\dag}\Psi^{(+)}\neq 0.
\end{equation}
The relation~(\ref{cond1}) now reads:
\begin{equation}
\label{3dlf}
\frac{2g\lambda}{1-\lambda}\partial^2
\left[\frac{1}{\partial^+}{\Psi^{(+)}}^{\dag}\Psi^{(+)}\right]=0.
\end{equation}
To meet the last equation one has to put $\lambda=0$.

Now we can take Eq.~(\ref{thetapmvph}) and set  $\lambda=0$,
$\Phi^+=0$. Then, using Eq.~(\ref{kindinlf}) we get the momentum
operator expressed through the independent fields $\Psi^{(+)}$,
$\bg{\Phi}^{\perp}$.
In Schr\"odinger representation we obtain
\begin{equation}
\label{pminintv}
\hat{P}_{\rho}^{int}=\omega_{\rho}\int d^4x\delta(\omega x)\,
H^{int}_{LC}(x),
\end{equation}
\begin{equation}
\label{hamiltonianv}
H^{int}_{LC}(x)=-g\bar{\psi}\sla{\varphi}\psi
+g^2\bar{\psi}\sla{\varphi}\frac{\sla{\omega}}{2i(\omega\cd\partial)}
\sla{\varphi}\psi
+g^2(\bar{\psi}\sla{\omega}\psi)\frac{1}{2(i\omega\cd\partial)^2}(\bar{\psi}
\sla{\omega}\psi).
\end{equation}
The first term in Eq.~(\ref{hamiltonianv}) is the standard spinor-vector
interaction Hamiltonian, while
the second and the third ones are the contact terms. Adding to this
Hamiltonian all necessary counterterms, one
gets Eq.~(\ref{hamiltonian_f}) with $H'(x)$ given by
Eq.~(\ref{hamlf}).


\section{Calculation of counterterms}\label{mo}

\subsection{Scalar boson}\label{scbos}

Explicit calculation of the fermion self-energy~(\ref{eq1sen}) is
similar to the one given in Ref.~\cite{dugne}. We will also calculate the
contribution~(\ref{sigfc}) of the fermion contact term. 
From Eq.~(\ref{eq1sen}) we can express the functions 
${\cal A}(p^2)$, ${\cal B}(p^2)$, and ${\cal C}(p^2)$ 
through $\Sigma(p)$:
\begin{equation}
\label{eq4sen} 
g^2{\cal A}(p^2)=\frac{1}{4}\mbox{Tr}[\Sigma(p)], \quad 
g^2{\cal B}(p^2)=\frac{m}{4\omega\cd p}
\mbox{Tr}[\Sigma(p)\sla{\omega }], \quad 
g^2{\cal C}(p^2)=\frac{1}{4m}\mbox{Tr}\left[\Sigma(p)\left(\sla{p}
-\frac{p^2\sla{\omega}}{\omega\cd p}\right)\right].
\end{equation}
For our purposes it is enough to know these function in the
region $p^2>0$ only.

We introduce the variables $k^+=k_0+k_z$, 
${\bf k}_\perp=(k_x,k_y)$ and $x=k^+/p^+$. For
convenience, we will carry out calculations in the frame where 
${\bf p}=0$
and $\omega=(1,0,0,-1)$. Hence,
$$
x=\frac{k^+ }{\sqrt{p^2}},\quad 
k_z=\frac{1}{2}k^+ -\frac{k_\perp^2+\mu^2}{2k^+},\quad
k_0=k^+ -k_z=\frac{1}{2}k^+ +\frac{k_\perp^2+\mu^2}{2k^+}.$$
Substituting the integral~(\ref{eq1sen}) for $\Sigma(p)$ in 
Eq.~(\ref{eq4sen}) 
[and similarly for Eq.~(\ref{sigfc})], we get
\begin{subequations}
\begin{eqnarray}
\label{abccalA}
{\cal A}(p^2)&=&-\frac{m}{16\pi^2}\int_0^{L^2} dk_\perp^2\int_0^1
dx\frac{1}{k_\perp^2+m^2x-p^2x(1-x)+\mu^2(1-x)},\\
\label{abccalB}
{\cal B}(p^2)&=&-\frac{m}{16\pi^2 }\int_0^{L^2} 
dk_\perp^2\int_0^1dx\frac{1-x}{k_\perp^2+m^2x-p^2x(1-x)+\mu^2(1-x)},\\
\label{abccalC}
{\cal C}(p^2)&=&-\frac{1}{32\pi^2 m}\int_0^{L^2} 
dk_\perp^2\int_\delta^{1-\delta}dx \frac{k_\perp^2+m^2-p^2(1-x)^2}
{(1-x)[k_\perp^2+m^2x-p^2x(1-x)+\mu^2(1-x)]},\\
\label{abccalCfc}
C_{fc}&=&\frac{1}{32\pi^2m }\int_0^{L^2} 
dk_\perp^2\left[\int_\delta^{1-\delta}\frac{dx } {x(1-x)}+
\int_{1+\delta}^{\infty}\frac{dx } {x(1-x)}\right].
\end{eqnarray}
\end{subequations}
We introduced in Eqs.~(\ref{abccalA})--(\ref{abccalCfc}) two cutoffs $L$ and $\delta$ in 
the variables $k_\perp^2$ and $x$, respectively, which restrict the integration
region as
$$
\delta <x<1-\delta,\,\,\,\,1+\delta<x<+\infty,\,\,\,\,k_{\perp}^2<L^2.
$$
In the following it is implied that $L\to\infty$ and $\delta\to 0$.
The two cutoffs do not constrain the integration in a spherically
symmetric domain and, in principle, violate rotational invariance. We will
see that $\delta m$ depends on the cutoff $L$ logarithmically and
does not depend on $\delta$.

In order to calculate the constants $A$, $B$, and $C$ we use
Eqs.~(\ref{abccalA})--(\ref{abccalC}) at $p^2=m^2$. We thus find
\begin{subequations}
\begin{eqnarray}
\label{eq7senA}
A&=&-\frac{m}{16\pi^2}\int_0^{L^2} dk_\perp^2\int_0^1
dx\frac{1}{k_\perp^2+m^2x^2+\mu^2(1-x)},\\
\label{eq7senB}
B&=&-\frac{m}{16\pi^2 }\int_0^{L^2}
dk_\perp^2\int_0^1dx\frac{1-x}{k_\perp^2+m^2x^2+\mu^2(1-x)},\\
\label{eq7senC}
C&=&-\frac{1}{32\pi^2 m}\int_0^{L^2}
dk_\perp^2\int_\delta^{1-\delta}dx \frac{k_\perp^2+m^2x(2-x)}
{(1-x)[k_\perp^2+m^2x^2+\mu^2(1-x)]}.
\end{eqnarray}
\end{subequations}
Calculating the integrals~(\ref{eq7senA})--(\ref{eq7senC}) for $A$, $B$, $C$, and 
those for $C_{fc}$ in Eq.~(\ref{abccalCfc}), and retaining only the terms
divergent at $L\to \infty$ and $\delta\to 0$, we get
\begin{subequations}
\begin{eqnarray}\label{Aa}
A  &=& -\frac{m}{16\pi^2} \log\frac{L^2}{m^2},\\
B&=&-\frac{m}{32\pi^2}\log\frac{L^2}{m^2},\label{Ab}\\
C&=&\hphantom{-}\frac{L^2\log\delta}{32\pi^2m}-
\frac{m}{32\pi^2}\log\frac{L^2}{m^2},\label{Ac}\\
C_{fc}&=&-\frac{L^2\log\delta}{32\pi^2m}.
\label{Afc}
\end{eqnarray}
\end{subequations}
From here we obtain
\begin{equation}
\label{cp}
Z_{\omega}=g^2(C+C_{fc})=-\frac{g^2m}{32\pi^2}\log\frac{L^2}{m^2}
\end{equation}
and
\begin{equation}
\label{dm}
\delta m= g^2(A+B+C+C_{fc})=-\frac{g^2m}{8\pi^2}\log\frac{L^2}{m^2}.
\end{equation}
Note that the strongest singularities [$\propto L^2$ and 
$\propto L^2\log\delta$]
cancel in the sum $C+C_{fc}$.

The normalization integral~(\ref{N}) reads
\begin{equation}\label{Ni}
N=4m^2a_1^2\left\{1+\frac{g^2}{16\pi^2}\int_0^{L^2} dk^2_{\perp}
\int_0^1 dx \frac{[k^2_{\perp}+m^2(x-2)^2]x}
{[k^2_{\perp}+m^2x^2+\mu^2(1-x)^2]^2}\right\}.
\end{equation}
The leading term of $N$ at $L\to\infty$ is
\begin{equation}\label{NN}
N=\frac{a_1^2 g^2 m^2}{8\pi^2}\log\frac{L^2m^6}{\mu^8}.
\end{equation}

Now let us prove that ${\cal C}(p^2)$ given by Eq.~(\ref{abccalC}) is a constant. 
For this purpose we calculate the difference
\begin{equation}\label{eqc3}
\Delta C(p^2)\equiv {\cal C}(p^2)-{\cal C}(m^2).
\end{equation}
Substituting $C(p^2)$ from Eq.~(\ref{abccalC}) into Eq.~(\ref{eqc3}), we obtain
\begin{equation}\label{eqc6}
\Delta C(p^2)=-\frac{g^2(p^2-m^2)}{32\pi^2 m}\int dk_\perp^2\int_0^1
\frac{[k_\perp^2(2x-1)+m^2x^2-\mu^2(1-x)^2]dx}
{[k_\perp^2+m^2x-(1-x)(p^2x-\mu^2)] [k_\perp^2+m^2x^2+\mu^2(1-x)]}.
\end{equation}
The integrand has no singularities at $x=0$ or $x=1$. Integrating over
$x$ results in zero for arbitrary $p^2$:
\begin{equation}\label{eqc7}
\Delta C(p^2)=0.
\end{equation}
This means that the $\omega$-dependent part, Eq.~(\ref{amplmom2}), 
of the fermion-boson scattering amplitude disappears, and the
whole amplitude, Eq.~(\ref{amplg41}), becomes $\omega$-independent.


\subsection{Gauge boson in the Feynman gauge}\label{gbosf}

The fermion self-energy with the gauge boson propagator in the Feynman
gauge is given by Eq.~(\ref{senf}). From this equation follows that 
the quantity $\Sigma^F_{\mu\nu}$ differs from the on-mass-shell self-energy
for the scalar boson case, $\Sigma$, defined by 
Eqs.~(\ref{eq8}) and~(\ref{eq1sen}), by the factor $-g_{\mu\nu}$ and by that
the gauge boson mass is zero. 
Hence, the coefficients $A_1$, $B_1$, and $C_1$ in Eq.~(\ref{selfenfeyn}) 
can be obtained directly from Eqs.~(\ref{Aa})--(\ref{Ac}) by
setting $\mu=0$, that does not make any influence on their asymptotics' at $L\to \infty$. 
Then, using the relations~(\ref{aa1}), we can easily find the coefficients $A$, $B$, and $C$ 
determining the self-energy~(\ref{senf}) in the Feynman gauge.
Therefore without repeating our previous calculations we obtain
\begin{subequations}
\begin{eqnarray}\label{Af}
A&=&\phantom{-}\frac{m}{4\pi^2}\log\frac{L^2}{m^2},\\
\label{Bf}
B&=&-\frac{m}{16\pi^2}\log\frac{L^2}{m^2},\\
\label{Cf}
C&=&\phantom{-}\frac{L^2\log\delta}{16\pi^2m}-\frac{m}{16\pi^2}\log\frac{L^2}{m^2},\\
\label{Cfc}
C_{fc}&=&-\frac{L^2\log\delta}{16\pi^2m}.
\end{eqnarray}
\end{subequations}
This gives
\begin{equation}\label{CpC}
Z_{\omega}=g^2(C+C_{fc})=-\frac{g^2m}{16\pi^2}\log\frac{L^2}{m^2},
\end{equation}
\begin{equation}\label{dmf}
\delta m= g^2(A+B+C+C_{fc})= \frac{g^2m}{8\pi^2}\log\frac{L^2}{m^2}.
\end{equation}
Like in the scalar boson case, 
the terms quadratically divergent in $L^2$ and
depending on the cutoff $\delta$ cancel in the sum $C+C_{fc}$.


\subsection{Gauge boson in the Light-Cone gauge}\label{gboslc}
The self-energy in the LC gauge is given by 
Eq.~(\ref{senflc}).
Doing the same as in deriving Eqs.~(\ref{eq7senA})--(\ref{eq7senC}), we obtain 
for the coefficients $A$, $B$, $C$ [and similarly for the contributions
from the fermion and boson contact terms $C_{fc}$ and $C_{bc}$
defined by Eqs.~(\ref{eq3b}) and~(\ref{eq3c}), respectively]:
\begin{subequations}
\begin{eqnarray}\label{eq4bA}
A&=&\frac{m}{8\pi^2}\int_0^{L^2} dk_\perp^2\int_0^1
dx\frac{1}{k_\perp^2+m^2x^2},\\
\label{eq4bB}
B&=&-\frac{m}{8\pi^2 }\int_0^{L^2} 
dk_\perp^2\int_0^1dx\frac{1-x}{k_\perp^2+m^2x^2},\\
\label{eq4bC}
C&=&-\frac{1}{16\pi^2 m}\int_0^{L^2} 
dk_\perp^2\int_\delta^{1-\delta}dx \frac{k_\perp^2(x^2-2x+2)+m^2x^3(2-x)}
{x^2(1-x)(k_\perp^2+m^2x^2)},\\
\label{eq4bCfc}
C_{fc}&=&\frac{1}{16\pi^2m }\int_0^{L^2} 
dk_\perp^2\left[\int_\delta^{1-\delta}\frac{dx } {x(1-x)}+
\int_{1+\delta}^{\infty}\frac{dx } {x(1-x)}\right],\\
\label{eq4bCbc}
C_{bc}&=&\frac{1}{16\pi^2m }\int_0^{L^2} 
dk_\perp^2\left[\int_0^{1-\delta}\frac{dx } {(1-x)^2}+
\int_{1+\delta}^{\infty}\frac{dx } {(1-x)^2}
-\int_0^{\infty}\frac{dx}{(1+x)^2}\right].
\end{eqnarray}
\end{subequations}
Calculating these
integrals, taking the limits $L\to \infty$ and $\delta\to 0$ and keeping
divergent terms only, we get
\begin{subequations}
\begin{eqnarray}\label{eq5bA}
A&=&\phantom{-}\frac{m}{8\pi^2}\log\frac{L^2}{m^2},\\
\label{eq5bB}
B&=&-\frac{m}{16\pi^2}\log\frac{L^2}{m^2},\\
\label{eq5bC}
C&=&-\frac{1}{\delta}\,\frac{L^2}{8\pi^2m}+
\frac{L^2\log\delta}{16\pi^2m}+\frac{L^2}{8\pi^2m}
+\frac{m}{16\pi^2}\log\frac{L^2}{m^2},\\
\label{eq5bCfc}
C_{fc}&=& - \frac{L^2\log\delta}{16\pi^2m},\\
\label{eq5bCbc}
C_{bc}&=&\hphantom{-}\frac{1}{\delta}\;\frac{L^2}{8\pi^2m}-\frac{L^2}{8\pi^2m}.
\end{eqnarray}
\end{subequations}
From here we find
\begin{equation}\label{cclc}
Z_{\omega}=g^2(C+C_{fc}+C_{bc})=\frac{m}{16\pi^2}\log\frac{L^2}{m^2}
\end{equation}
and
\begin{equation}\label{eq6b}
\delta m=
g^2(A+B+C+C_{fc}+C_{bc})=\frac{g^2m}{8\pi^2}\log\frac{L^2}{m^2}.
\end{equation}
The terms $C$, $C_{fc}$, and $C_{bc}$ are quadratically and
logarithmically divergent in the cutoff $L$ and divergent in the cutoff
$\delta$ like $1/\delta$ and $\log\delta$. 
However, all the divergences excepting $\log(L^2/m^2)$ ones 
cancel in the sum $C+C_{fc}+C_{bc}$.

\section{Equations for the spin components}

\subsection{Scalar boson}\label{eqscb}

The spin components $a_1$, $b_1$, and  $b_2$
in the scalar boson case are defined by
Eqs.~(\ref{eqf3}) and~(\ref{eq7}).
As explained in Sec.~\ref{solut1}, the equations for the spin
components are obtained by substituting
these equations into Eqs.~(\ref{eq4}) and~(\ref{eq5}) and
taking
traces.
As a result, we get the following system of three homogeneous equations
for the three unknown constants $a_1$, $b_1$, and  $b_2$:
\begin{multline}\label{eq_g10}
2a_1m({\delta m'}^2 - 2m\delta m' - 2m Z)-
b_1g[(A+B)(\delta m'-2m)-2 m C]\\
+2b_2g[(A+B)m-\delta m' B]=0,
\end{multline}
\begin{multline}
\label{eq_g2a0}
2a_1gm[m^2(3+x)-m\delta m'(1+x)+s(1-x)]+b_1[g^2m(A+B)(1+x)-m^2(3+x)-s(1-x)]\\
+2b_2m(g^2B-m)(1+x)=0,
\end{multline}
\begin{equation}
\label{eq_g2b0}
2a_1gm[m(1+x)-\delta m'x]+b_1[g^2(A+B)x-m(1+x)]+2b_2(g^2B-m)x=0.
\end{equation}
The condition that the determinant~(\ref{detsc}) of this system is zero results in the
solution~(\ref{eq12}),~(\ref{eq12a}).


\subsection{Feynman gauge}\label{eqsfb}

In the case of gauge boson in the Feynman gauge, the one-body fermion
vertex function $\Gamma_1$ is determined by one spin component only, as given
by Eq.~(\ref{gamma1feyn}), while
the two-body vertex function $\Gamma_2$ by eight spin components, four of them
being zeroes when sectors higher than $|Fb\rangle$ are omitted in
calculations.
The derivation of the system of equations for the spin components is
described
in Sec.~\ref{solut2}. The system of five equations for $a_1$, $b_{1-4}$ is
given by
\begin{multline}
a_1m(2m\delta m'-{\delta m'}^2+2mZ)-
b_1g[(2A_1-B_1)(\delta m'-2m)+2C_1m] \\
\label{eqF1feyn}
+b_2mg(A_1+B_1)-2b_3g[A_1m+B_1(\delta m'-2m)]+2b_4gB_1m=0,
\end{multline}
\begin{multline}\label{eq1feyn}
2a_1gm\left[m^2(3-x)-s(1-x)-m\delta m'(2-x)\right] \\
-b_1\left[2mg^2(2A_1-B_1)(2-x)+m^2(3-x)-s(1-x)\right] \\
-b_2m^2(1+x)+2b_3m[2g^2B_1(x-2)+m(1-2x)]-2b_4m^2x=0,
\end{multline}
\begin{equation}
\label{eq2feyn}
(2a_1gm-b_1)(1+x)=0,
\end{equation}
\begin{equation}
\label{eq3feyn}
   2a_1gm[x\delta m'+m(1-2x)]+b_1[2g^2(2A_1-B_1)x-m(1-2x)]+2b_3(2g^2B_1-m)x=0,
\end{equation}
\begin{multline}\label{eq4feyn}
2a_1g\left\{m^3(1+x)+[2m^2+s(1-x)](m-\delta m')\right\}\\
-b_1\left\{m^2(1+x)+\frac{[2m^2+s(1-x)][2g^2
(2A_1-B_1)+m]}{m}\right\}\\
-b_2[m^2(3+x)+s(1-x)]-2b_3\left[m^2(1+x)+2g^2B_1
\frac{2m^2+s(1-x)}{m}\right]\\
-2b_4m^2(1+x)=0.
\end{multline}
The determinant of this system of equations is given by
Eq.~(\ref{determfeyn}). The condition $Det=0$ results in the
solution~(\ref{delmfeyn})--(\ref{b2feyn}).


\subsection{Light-Cone gauge}\label{eqslc}

The equations for the spin components, derived in
Sec.~\ref{solut3}, have the form
\begin{multline}\label{eqF1lf}
a_1m[2m\delta m'-{\delta m'}^2+2mZ]+
b_1g[\delta m'(B_1-A_1)+2m(A_1-2B_1-B_2+5C_1+4C_1')]\\
+2b_2g[m(B_1-A_1)-B_1\delta m']=0,
\end{multline}
\begin{multline}\label{eq1lf}
2a_1gm\left[\frac{1+x^2}{1-x}(s-m^2)-2m^2+m\delta m'(1-x)\right]\\
-b_1\left[\frac{1+x^2}{1-x}(s-m^2)-2m^2-2g^2m(A_1-B_1)(1-x)\right]\\
+2b_2m(2g^2B_1-m)(1-x)=0,
\end{multline}
\begin{equation}
\label{eq2lf}
2a_1gm[m(1-x)+x\delta m']-b_1[m(1-x)-2g^2(A_1-B_1)x]+2b_2(2g^2B_1-m)x=0.
\end{equation}
The condition that the determinant of this system is zero [see Eq.~(\ref{eq1lfb})] 
results in the solution~(\ref{delmlf})--(\ref{b2lf}).

\section{Dependence of $\delta m$ on the type of cutoff}\label{cutoffs}

Here we will show how to adjust the cutoffs in Eqs.~(\ref{eq4bA})--(\ref{eq4bCbc}) 
in order to 
reproduce the perturbative value of $\delta m$ obtained in Ref.~\cite{must91} 
for the LC gauge.

\begin{figure}[btph]
\begin{center}
\centerline{\epsfbox{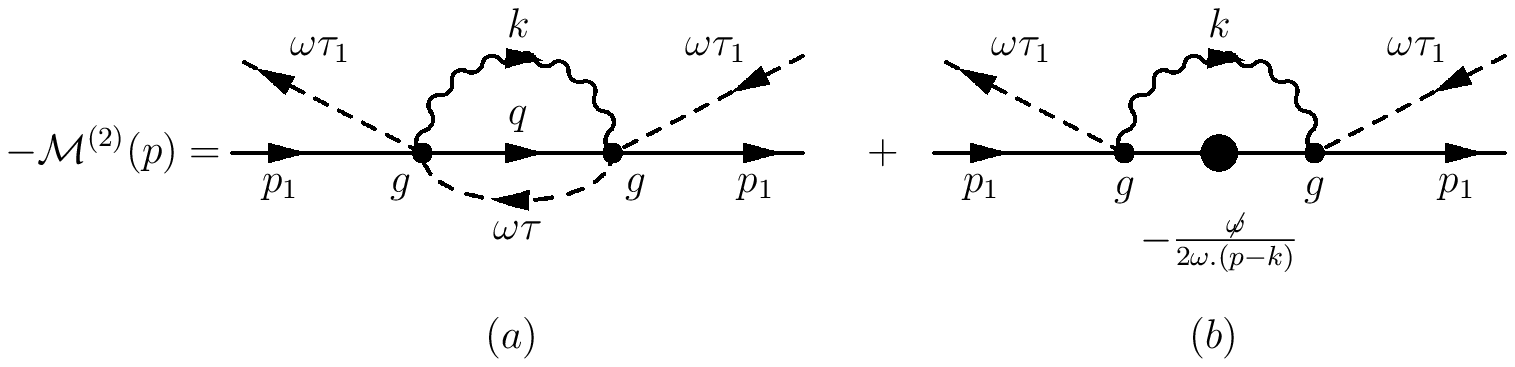}}
\caption{Mass operator in CLFD. The lowest order contributions are shown:
the light-front self-energy, $-\Sigma(p)$ (a), and the contact term,
$-\Sigma_{fc}(p)$ (b). Here $p=p_1-\omega\tau_1$.}\label{fig0a}
\end{center}
\end{figure}

\begin{figure}[btph]
\begin{center}
\centerline{\epsfbox{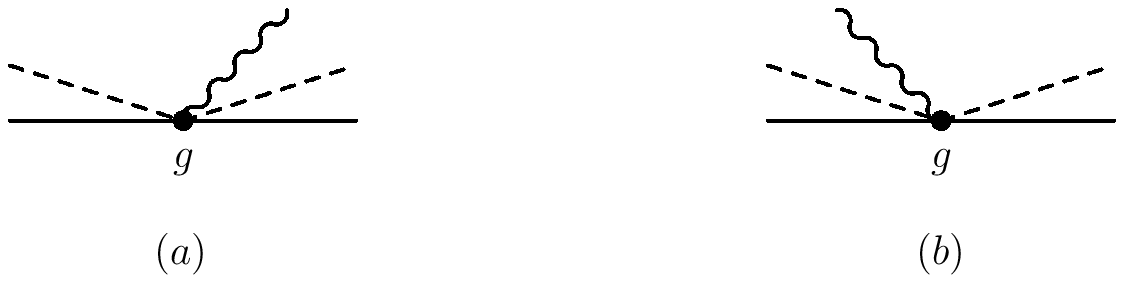}}
\caption{Interaction vertices generated by the Hamiltonian $H_1(x)$,
Eq.~(\protect{\ref{h1}}).\label{fig1}}
\end{center}
\end{figure}

\begin{figure}[btph]
\begin{center}
\centerline{\epsfbox{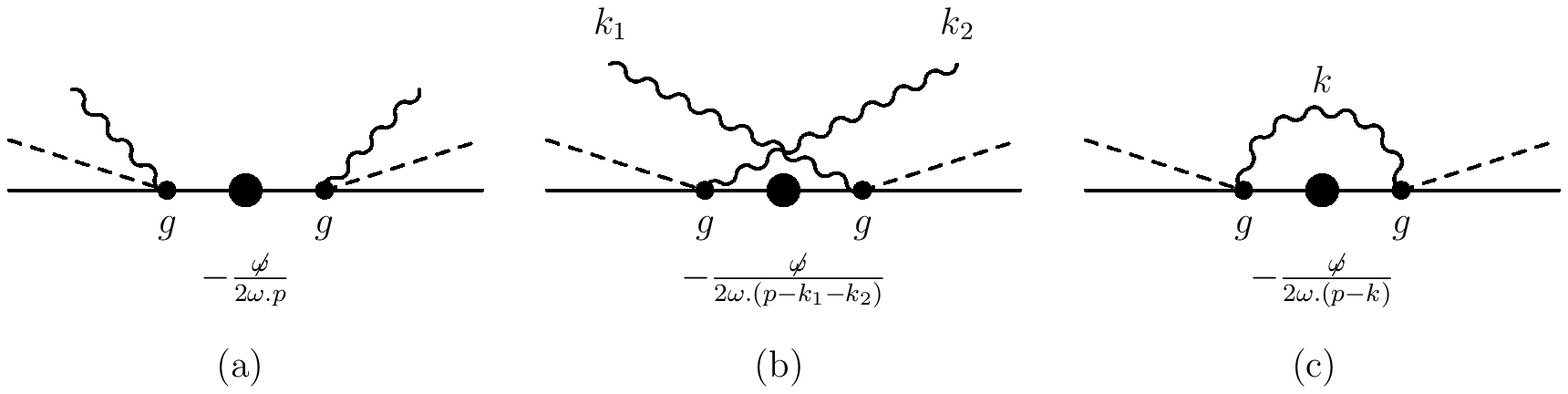}}
\caption{Interaction vertices generated by the Hamiltonian $H_{1c}(x)$,
Eq.~(\protect{\ref{h1c}}).\label{fig2}}
\end{center}
\end{figure}

\begin{figure}[btph]
\begin{center}
\centerline{\epsfbox{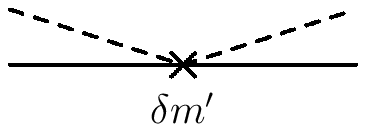}}
\caption{Interaction vertex generated by the Hamiltonian $H_{2}(x)$,
Eq.~(\protect{\ref{h2c}}).\label{fig3}}
\end{center}
\end{figure}

\begin{figure}[btph]
\begin{center}
\centerline{\epsfbox{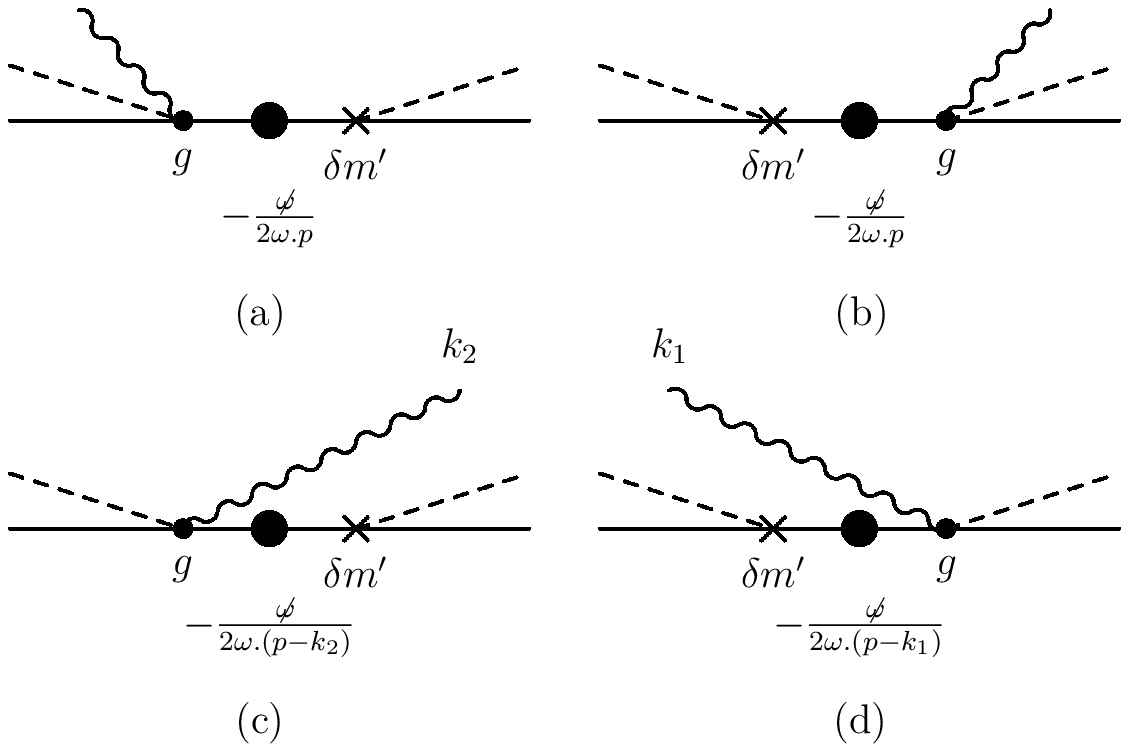}}
\caption{Interaction vertices generated by the Hamiltonian $H_{2c}(x)$,
Eq.~(\protect{\ref{h3c}}).\label{fig4}}
\end{center}
\end{figure}

\begin{figure}[btph]
\begin{center}
\centerline{\epsfbox{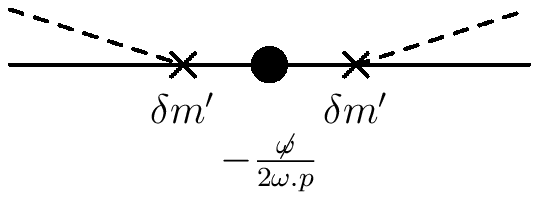}}
\caption{Interaction vertex generated by the Hamiltonian
$H^\prime_{2c}(x)$,
Eq.~(\protect{\ref{h4c}}).\label{fig5}}
\end{center}
\end{figure}

\begin{figure}[btph]
\begin{center}
\centerline{\epsfbox{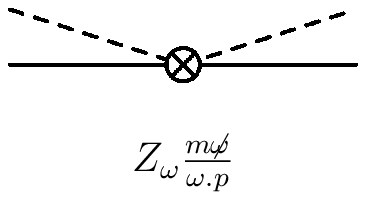}}
\caption{Interaction vertex generated by the Hamiltonian $H_{3}(x)$,
Eq.~(\protect{\ref{h5c}}).\label{fig6}}
\end{center}
\end{figure}

\begin{figure}[btph]
\begin{center}
\centerline{\epsfbox{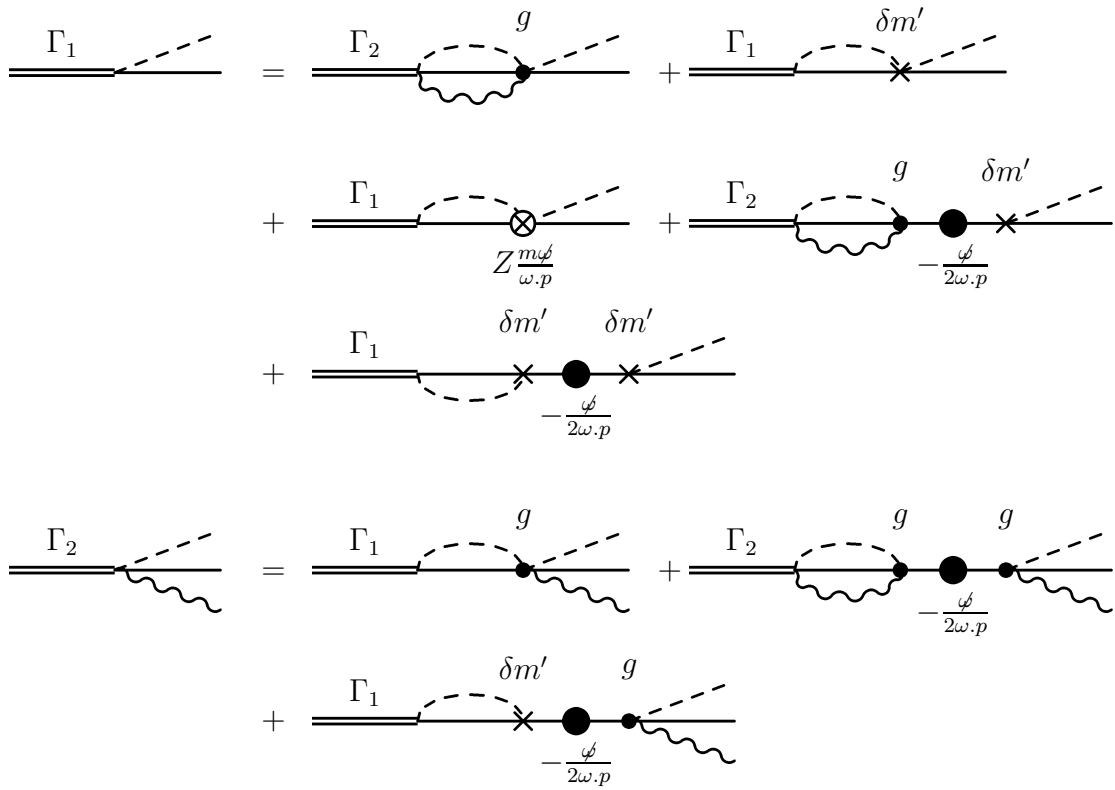}}
\caption{System of equations for the one- and two-body vertex functions
for the scalar boson case.
\label{fig4a}}
\end{center}
\end{figure}

\begin{figure}[btph]
\begin{center}
\centerline{\epsfbox{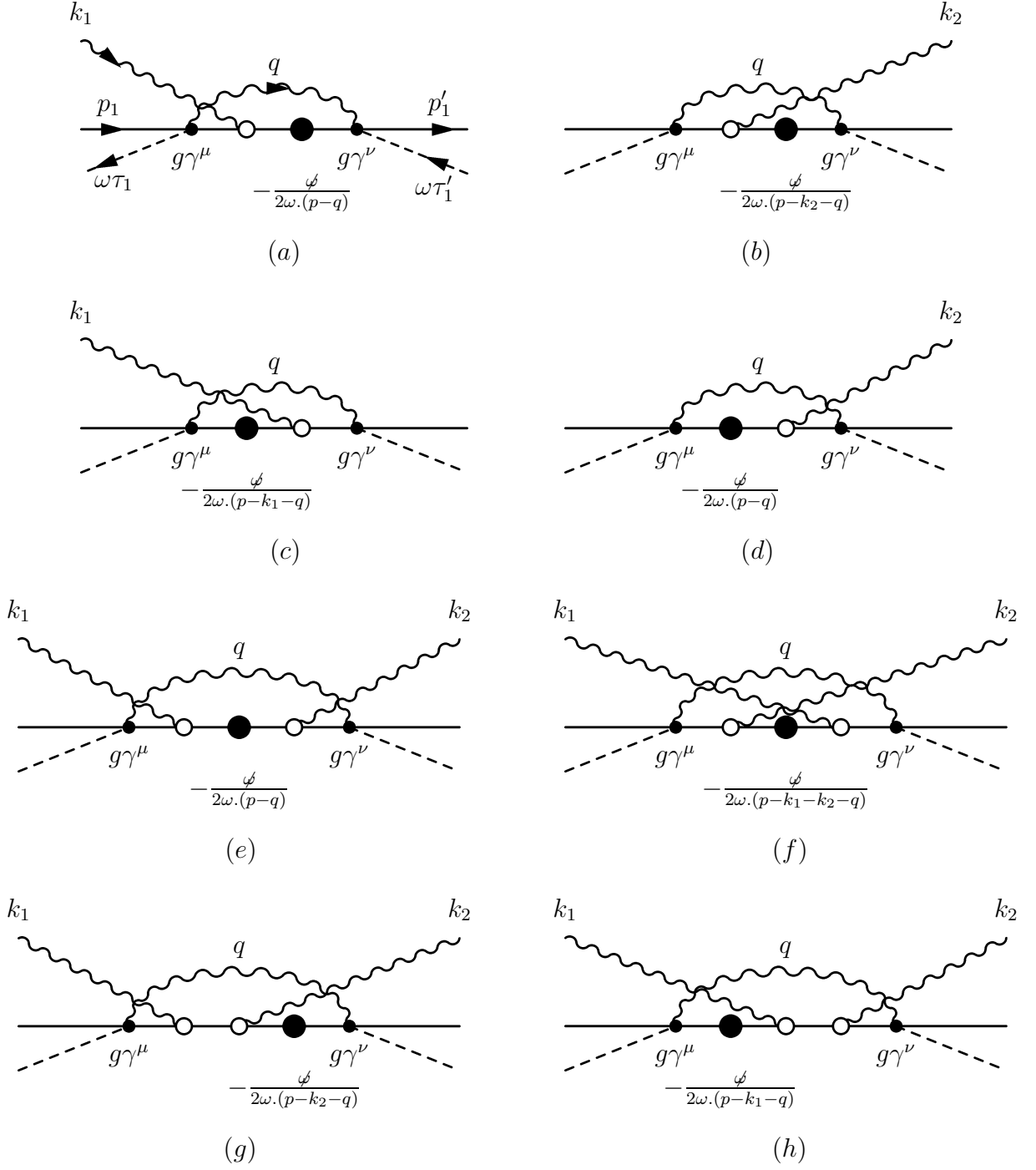}}
\caption{Interaction vertices generated
in the Feynman gauge by the
piece $H_{1F}'(x)$, Eq.~(\protect{\ref{hamfeyn1}}), of the light-front
Hamiltonian.
\label{figmelf}}
\end{center}
\end{figure}

\begin{figure}[btph]
\begin{center}
\centerline{\epsfbox{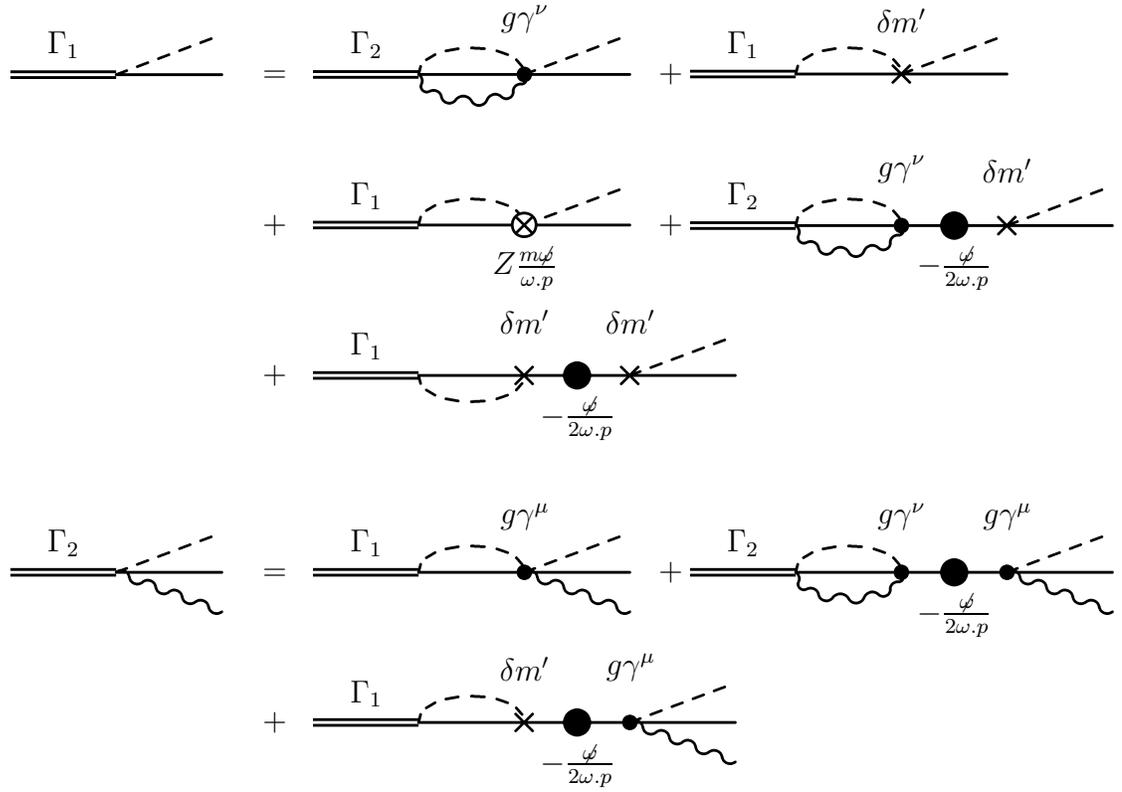}}
\caption{System of equations for the one- and two-body vertex functions
for the case of vector boson in the Feynman gauge.\label{figFeyn}}
\end{center}
\end{figure}

\begin{figure}[btp]
\begin{center}
\centerline{\epsfbox{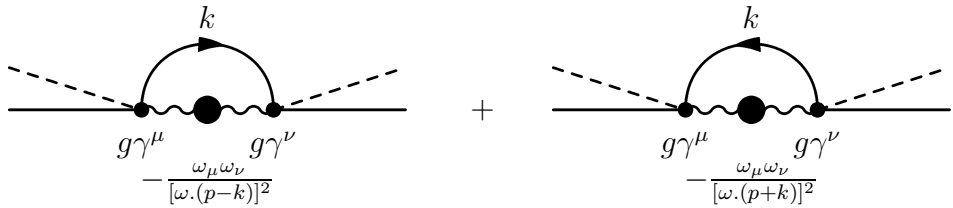}} \caption{Interaction vertices
generated in the LC gauge by the piece $H_{LC}'(x)$,
Eq.~(\protect{\ref{hamlf}}), of the light-front Hamiltonian. \label{fig6a}}
\end{center}
\end{figure}

The coefficients $A$, $B$, and $C$ proceed from the light-front
diagram shown in Fig.~\ref{fig0a}(a), 
whereas $C_{fc}$ and $C_{bc}$ originate from
the loop diagrams shown in Figs.~\ref{fig0a}(b) 
and~\ref{fig6a}, respectively. Following
Ref.~\cite{must91}, we regularize the amplitude of the diagram
of Fig.~\ref{fig0a}(a) by the condition
\begin{equation}
\label{apd1}
s=\frac{k_{\perp}^2}{x}+\frac{k_{\perp}^2+m^2}{1-x}<\Lambda^2
\end{equation} 
with the additional requirement
\begin{equation}
\label{apd2}
k_{\perp}<L\ll \Lambda.
\end{equation} 
Eq.~(\ref{apd1}) leads to the restrictions $\alpha<x<1-\beta$, where
\begin{equation}
\label{apd3}
\alpha=\frac{k_{\perp}^2}{\Lambda^2},\,\,\,\,
\beta=\frac{k_{\perp}^2+m^2}{\Lambda^2}. 
\end{equation}
The integrands in the expressions for $A$ and $B$ 
[see Eqs.~(\ref{eq4bA}), (\ref{eq4bB})] are not singular in $x$, hence, 
$A$ and $B$ are given by the same formulas~(\ref{eq5bA}),~(\ref{eq5bB}).
The expression~(\ref{eq4bC}) transforms to
$$
C=-\frac{1}{16\pi^2m}\int_{0}^{L^2}dk_{\perp}^2
\int_{\alpha}^{1-\beta}
dx\,\frac{k_{\perp}^2(x^2-2x+2)+m^2x^3(2-x)}{x^2(1-x)(k_{\perp}^2+m^2x^2)}
$$
$$
=\frac{1}{16\pi^2m}\int_0^{L^2}dk_{\perp}^2
\left[\int_0^1
\frac{2m^2(1-x)dx}{m^2x^2+k_{\perp}^2}-\int_{\alpha}^1
\frac{2dx}{x^2}-\int_0^{1-\beta}\frac{dx}{1-x}
\right]
$$
\begin{equation}
\label{apd4}
=\frac{m}{16\pi^2}\log\frac{L^2}{m^2}+
\frac{1}{16\pi^2m}\int_0^{L^2}dk_{\perp}^2
\left[
\log\beta-\frac{2}{\alpha}+2\right].
\end{equation} 
As far as the divergent integrals over $x$ for the coefficients 
$C_{fc}$ and $C_{bc}$ in Eqs.~(\ref{eq4bCfc}), (\ref{eq4bCbc})
are concerned, the prescription of Ref.~\cite{must91} demands to use for 
their regularization the only cutoff $\alpha$.
Therefore, we have
\begin{equation}
\label{apd5}
C_{fc}=\frac{1}{16\pi^2m}\int_{0}^{L^2}dk_{\perp}^2
\left[
\int_{\alpha}^{1-\alpha}\frac{dx}{x(1-x)}+
\int_{1+\alpha}^{\infty}\frac{dx}{x(1-x)}
\right]=-\frac{1}{16\pi^2m}
\int_{0}^{L^2}dk_{\perp}^2\,\log\alpha,
\end{equation}
$$
C_{bc}=\frac{1}{16\pi^2m}\int_{0}^{L^2}dk_{\perp}^2
\left[
\int_{0}^{1-\alpha}\frac{dx}{(1-x)^2}+
\int_{1+\alpha}^{\infty}\frac{dx}{(1-x)^2}-
\int_{0}^{\infty}\frac{dx}{(1+x)^2}
\right]
$$
\begin{equation}
\label{apd6}
=\frac{1}{16\pi^2m}
\int_{0}^{L^2}dk_{\perp}^2\left[
\frac{2}{\alpha}-2
\right].
\end{equation}
Now, using the first of Eqs.~(\ref{soldelmlc}) we get
\begin{equation}
\label{apd7}
\delta m=\frac{g^2m}{8\pi^2}\log\frac{L^2}{m^2}
+\frac{g^2}{16\pi^2m}
\int_{0}^{L^2}dk_{\perp}^2\log
\frac{\beta}{\alpha}.
\end{equation}
Taking into account Eqs.~(\ref{apd3}) and performing the integration
we finally obtain
\begin{equation}
\label{apd8}
\delta m=\frac{3g^2m}{16\pi^2}\log\frac{L^2}{m^2}.
\end{equation}
The same coefficient $3/16$ as in Eq.~(\ref{apd8}) appears also for
the regularized perturbative $\delta m_{4DF}$ in standard 
4DF approach:
\begin{equation}
\label{apd9}
\delta m_{4DF}=\frac{3g^2m}{16\pi^2}\log\frac{L_{4DF}^2}{m^2},
\end{equation}
where $L_{4DF}$ is a relativistically invariant cutoff restricting
the region of integration in four-dimensional space.
Note however that in spite of the similarity between Eqs.~(\ref{apd8})
and~(\ref{apd9}) the cutoffs $L$ and $L_{4DF}$
restrict different integration regions and therefore have different meaning.

So, the only difference between our method of calculation and the 
method from Ref.~\cite{must91}
consists in that integrating over $x$ we apply the same cutoff $\delta$ 
for all the divergent integrals in Eqs.~(\ref{eq4bC})--(\ref{eq4bCbc}), while the authors 
of Ref.~\cite{must91} used two different cutoffs, $\alpha$ and $\beta$
(dependent on $k_{\perp}^2$), 
as exposed above. Putting $\alpha=\beta$ in Eq.~(\ref{apd7}) immediately 
reproduces our result~(\ref{eq6b}) for $\delta m$.

\newpage


\end{document}